\documentclass[11pt,a4paper,twoside]{article}
\usepackage{dina4h}
\usepackage{epsfig,citesort,rotating,axodraw}

\def\3{\ss} 

\sloppypar

\newcommand{\keV}{\,\mathrm{ke\kern-1pt V}}
\newcommand{\MeV}{\,\mathrm{Me\kern-1pt V}}
\newcommand{\GeV}{\,\mathrm{Ge\kern-1pt V}}
\newcommand{\EGeV}{E\,\mathrm{[Ge\kern-1pt V]}}
\newcommand{\WGeV}{W\,\mathrm{[Ge\kern-1pt V]}}
\newcommand{\TeV}{\,\mathrm{Te\kern-1pt V}}
\newcommand{\cm}{\,\mathrm{cm}} 

\newcommand{\mm}{\,\mathrm{mm}}

\newcommand{\mum}{\,\mu\mathrm{m}}

\newcommand{\m}{\,\mathrm{m}}

\newcommand{\mrad}{\,\mathrm{mrad}}
 
\newcommand{\pom}{{\rm I\! P}}
\newcommand{\reg}{{\rm I\! R}}
\newcommand{\sgsp}{\sigma_\mathrm{tot}^{\gamma^*p}}

\newcommand{\ft}{F_2} 
\newcommand{\ftxq}{F_2(x,Q^2)}
\newcommand{\fl}{F_L}

\newcommand{\emax}{\eta_\mathrm{max}}

\newcommand{\pb}{\,\mathrm{pb}^{-1}}
\newcommand{\mub}{\,\mu\mathrm{b}}

\newcommand{\zpc}[1]{Z.\ Phys.\ {\bf C #1}} 

\newcommand{\prl}[1]{Phys.\ Rev.\ Lett.\ {\bf #1}}
\newcommand{\prd}[1]{Phys.\ Rev.\ {\bf D #1}}

\newcommand{\plb}[1]{Phys.\ Lett.\ {\bf B #1}}
\newcommand{\npb}[1]{Nucl.\ Phys.\ {\bf B #1}}
\newcommand{\npbp}[1]{Nucl.\ Phys.\ {\bf B} (Proc.\ Suppl.) {\bf #1}}
\newcommand{\epjc}[1]{Eur.\ Phys.\ J.\ {\bf C #1}}
\newcommand{\cpc}[1]{Comp.\ Phys.\ Comm. {\bf #1}}

\newcommand{\nima}[1]{Nucl.\ Instr.\ Methods {\bf A #1}}

\newcommand{\dy}{DESY Report } 
\newcommand{\ce}{CERN Report }

\begin{document} \thispagestyle{empty} 

\title{{\tt\normalsize\hspace*{28pt}DESY 00-071\hfill ISSN
0418-9833\hspace*{28pt} \\\hspace*{28pt}May 2000\hfill\
\hspace*{28pt} \\\hspace*{28pt}Revised Version\hfill\
\hspace*{28pt}\\}\vskip2cm \bf\LARGE\boldmath Measurement of the\\
\bf\LARGE Proton Structure Function $\ft$\\ \bf\LARGE at Very Low
$Q^2$ at HERA\\ \vskip2cm }

\author{\Large ZEUS Collaboration\vspace{1cm}}

\date{}

\maketitle

\vfill \centerline{\bf Abstract} \vskip4.mm \centerline{
\begin{minipage}{15.cm} 
\noindent A measurement of the proton structure function $\ft(x,Q^2)$
is presented in the kinematic range $0.045\GeV^2<Q^2<0.65\GeV^2$ and
$6\cdot 10^{-7}<x<1\cdot 10^{-3}$. The results were obtained using a
data sample corresponding to an integrated luminosity of $3.9\pb$ in
$e^+p$ reactions recorded with the ZEUS detector at
HERA\@. Information from a silicon-strip tracking detector, installed
in front of the small electromagnetic calorimeter used to measure the
energy of the final-state positron at small scattering angles,
together with an enhanced simulation of the hadronic final state, has
permitted the extension of the kinematic range beyond that of previous
measurements. The uncertainties in $\ft$ are typically less than
4\%. At the low $Q^2$ values of the present measurement, the rise of
$\ft$ at low $x$ is slower than observed in HERA data at higher $Q^2$
and can be described by Regge theory with a constant logarithmic
slope $\partial{\ln}F_2/\partial{\ln}(1/x)$.
The dependence of $\ft$ on $Q^2$ is stronger than at higher
$Q^2$ values, approaching, at the lowest $Q^2$ values of this
measurement, a region where $\ft$ becomes nearly proportional to
$Q^2$.
\end{minipage} } \vfill

\thispagestyle{empty} \newpage

%
%
%
%

\pagenumbering{Roman} 
\begin{center} 
{ \Large The ZEUS Collaboration } 
\end{center} 
 J.~Breitweg, 
 S.~Chekanov, 
 M.~Derrick, 
 D.~Krakauer, 
 S.~Magill, 
 B.~Musgrave, 
 A.~Pellegrino, 
 J.~Repond, 
 R.~Stanek, 
 R.~Yoshida\\ 
 {\it Argonne National Laboratory, Argonne, IL, USA}~$^{p}$
% ...
\newpage 
% ...
                                               %
                                               %
\newpage                                       
                                               %
                                               %
\begin{tabular}[h]{rp{14cm}} 
$^{a}$ & supported by the Natural Sciences and Engineering Research 
 Council of Canada (NSERC) \\ 
$^{b}$ & supported by the FCAR of Qu\'ebec, Canada \\ 
$^{c}$ & supported by the German Federal Ministry for Education and 
 Science, Research and Technology (BMBF), under contract 
 numbers 057BN19P, 057FR19P, 057HH19P, 057HH29P, 057SI75I \\ 
$^{d}$ & supported by the MINERVA Gesellschaft f\"ur Forschung GmbH, the 
German Israeli Foundation, the Israel Science Foundation, the Israel 
Ministry of Science and the Benozyio Center for High Energy Physics\\ 
$^{e}$ & supported by the German-Israeli Foundation, the Israel Science 
 Foundation, the U.S.-Israel Binational Science Foundation, and by 
 the Israel Ministry of Science \\ 
$^{f}$ & supported by the Italian National Institute for Nuclear Physics 
 (INFN) \\ 
$^{g}$ & supported by the Japanese Ministry of Education, Science and 
 Culture (the Monbusho) and its grants for Scientific Research \\ 
$^{h}$ & supported by the Korean Ministry of Education and Korea Science 
 and Engineering Foundation \\ 
$^{i}$ & supported by the Netherlands Foundation for Research on 
 Matter (FOM) \\ 
$^{j}$ & supported by the Polish State Committee for Scientific Research, 
 grant No. 112/E-356/SPUB/DESY/P03/DZ 3/99, 620/E-77/SPUB/DESY/P-03/ 
 DZ 1/99, 2P03B03216, 2P03B04616, 2P03B03517, and by the German 
 Federal Ministry of Education and Science, Research and Technology (BMBF)\\ 
$^{k}$ & supported by the Polish State Committee for Scientific 
 Research (grant No. 2P03B08614 and 2P03B06116) \\ 
$^{l}$ & partially supported by the German Federal Ministry for 
 Education and Science, Research and Technology (BMBF) \\ 
$^{m}$ & supported by the Fund for Fundamental Research of Russian Ministry 
 for Science and Edu\-cation and by the German Federal Ministry for 
 Education and Science, Research and Technology (BMBF) \\ 
$^{n}$ & supported by the Spanish Ministry of Education 
 and Science through funds provided by CICYT \\ 
$^{o}$ & supported by the Particle Physics and 
 Astronomy Research Council \\ 
$^{p}$ & supported by the US Department of Energy \\ 
$^{q}$ & supported by the US National Science Foundation 
\end{tabular} 
                                               %


\newpage
\parindent0.6cm \parskip0.0cm
\pagenumbering{arabic}

\section{Introduction}

A remarkable feature of the proton structure function $\ftxq$ is its
rapid rise at low $x$, observed by the H1 and ZEUS collaborations
\cite{ZEUS+H194} at HERA\@. First observed for $Q^2$ values above
$10\GeV^2$, the persistence of this rise down to small $Q^2$
\cite{ZEUSBPC,H1SVX95,ZEUSSVX95} challenges our understanding of
QCD. In a recent publication \cite{ZEUSSVX95}, the ZEUS collaboration
has discussed the transition from deep inelastic scattering to
photoproduction. It was found that standard non-perturbative
approaches which apply in photoproduction fail to describe the data in
the region above $Q^2=0.9\GeV^2$. Next-to-leading-order QCD fits are
successful when taken down to these low $Q^2$ values. As $Q^2$
approaches $1\GeV^2$, however, these fits yield vanishing gluon
densities at low $x$, while the sea-quark density remains finite. Such
a ``valence-like'' gluon distribution, vanishing as $x \rightarrow 0$,
seems unnatural even at low $Q^2$ and has led to much discussion
\cite{ZEROGLUON}. Precise measurements at low $Q^2$ are important in
elucidating this subject.

This letter presents a measurement of $\ft$ at low $Q^2$
($0.045\GeV^2<Q^2<0.65\GeV^2$) and low $x$ ($6\cdot 10^{-7}<x<1\cdot
10^{-3}$). The data used correspond to an integrated luminosity of
$3.9\pb$ and were taken with dedicated triggers during six weeks of
$e^+p$ running in 1997. Compared to a previous result\footnote{When
the previous measurement of $\ft$ at low $Q^2$ was published, it was
found that existing models and parameterizations
\cite{OLDER_PARAMETERIZATIONS} were unable to describe the
data. Subsequently, there have been various new approaches, as well as
updates and improvements of existing ones, to describe $\ftxq$ in the
entire $Q^2$ range \cite{ALLM97,NEWER_PARAMETERIZATIONS}. The validity
of these recent models is not discussed here.} \cite{ZEUSBPC}, the new
measurement covers a larger kinematic region with an improved
statistical precision and systematic accuracy. This was made possible
by the addition of a Beam Pipe Tracker in front of the Beam Pipe
Calorimeter used for measuring the energy of the final-state positron
at small scattering angles, and by an enhanced simulation of the
hadronic final state.

\section{Kinematic variables and cross sections}

Inclusive deep inelastic positron-proton scattering, $e^+p\to e^+X$,
can be described in terms of two kinematic variables, $x$ and $Q^2$,
where $x$ is the Bjorken scaling variable and $Q^2$ the negative of
the square of the four-momentum transfer. They are defined as
$Q^2=-q^2=-(k-k')^2$ and $x=Q^2/(2P\cdot q)$, where $k$ and $P$ are
the four-momenta of the incoming positron and proton, respectively,
and $k'$ is the four-momentum of the scattered positron. The
fractional energy transferred to the proton in its rest frame, $y$, is
related to $x$ and $Q^2$ by $Q^2=sxy$, where $s=4E_eE_p$ is the square
of the positron-proton center-of-mass energy. Here, $E_e = 27.5\GeV$
and $E_p = 820\GeV$ are the positron and proton beam energies,
respectively.

At the low $Q^2$ values of this measurement, where the contribution
from $Z$ exchange is negligible, the double-differential cross section
for inelastic $e^{+}p$ scattering can be written in terms of $\ft$ and
the longitudinal structure function $\fl$ as
\begin{equation}\frac{d^2\sigma}{dx\,dQ^2}=\frac{2\pi\alpha^2}{xQ^4} 
\left((1+(1-y)^2)\ft-y^2\fl\right)(1+\delta_r).\end{equation} The QED
radiative correction, $\delta_r$, is a function of $x$ and $Q^2$, but,
to a good approximation, independent of both $\ft$ and $\fl$ \cite{RADCORR}.

\section{Experimental setup and kinematic reconstruction}

The ZEUS detector has been described in detail previously
\cite{ZEUSDET}. In the present analysis, the scattered positron was
detected in the Beam Pipe Calorimeter (BPC) and Beam Pipe Tracker
(BPT) \cite{BPCTEXP,CATHESIS}. The BPC was installed in 1995 to
enhance the acceptance of the ZEUS detector for low-$Q^2$ events,
where the positron is scattered through a small angle, and was used
for a previous measurement of $\ft$ \cite{ZEUSBPC}. In 1997, the BPT
was installed in front of the BPC to complement the calorimetric
energy measurement with precise tracking information.

\subsection{Beam Pipe Calorimeter and Beam Pipe Tracker}

The BPC is a tungsten-scintillator sampling calorimeter with the front
face located at $Z=-293.7\cm$, the center at $Y=0.0\cm$, and the inner
edge of the active area at $X=4.4\cm$, as close as possible to the
rear beam pipe\footnote{The ZEUS right-handed Cartesian coordinate
system has its origin at the nominal interaction point, the $Z$ axis
pointing in the proton beam direction (referred to as forward
direction), and the $X$ axis pointing toward the center of HERA\@. The
polar angle $\vartheta$ is measured with respect to the positive $Z$
axis; it is convenient to also define $\theta=\pi-\vartheta$. The
pseudorapidity is defined as $\eta=-\ln(\tan(\vartheta/2))$.}. The BPC
has an active area of $12.0 \times 12.8\cm^2$ in $X\times Y$ and its
depth in $Z$ corresponds to $24\,X_0$. The relative energy resolution
as determined in test-beam measurements with 1--6$\GeV$ electrons is
${\Delta E}/{E}={17\%}/{\sqrt{\EGeV}}$. The scintillator layers are
alternately subdivided into vertical and horizontal strips, $7.9\mm$
in width, referred to as $X$ and $Y$ fingers, respectively. The stacks
of fingers with common $X$ and $Y$ positions are read out together by
a wavelength shifter bar.

The BPT is a silicon microstrip tracking device. In 1997, it was
equipped with two detector planes to measure the $X$ coordinate. The
planes, of dimensions $6\times 6\times 0.03\cm^3$, have 576 strips
with a $100\mum$ pitch. The detectors are read out with binary
electronics \cite{LPSSCHEME}. The number of readout channels per
detector is 512; 128 strips at the outer edge of a plane are read out
in pairs, thus giving an effective pitch of $200\mum$. The detector
planes and front-end electronics are supported in a carbon-fiber
composite structure attached to the front face of the BPC\@. The
planes are positioned at $Z=-252.7\cm$ and $Z=-279.1\cm$ and have
centers at $X=6.85\cm$ and $Y=0.0\cm$.

Scattered positrons traveling from the interaction point toward the
BPC and BPT leave the beam pipe through a $1.5\mm$ thick
($1.6\%\,X_0$) aluminum exit window positioned at $Z=-249.8\cm$. The
fiducial area of the measurement was defined by the overlap of the
exit window with the acceptance of BPC and BPT, and was thus limited
to a D-shaped region in the range $5.2\cm<X<9.3\cm$ and
$-2.3\cm<Y<2.8\cm$ at $Z=-293.7\cm$. Only positrons scattered into the
fiducial area were used in the analysis, yielding an angular
acceptance of $18\mrad<\theta_e<32\mrad$ for particles emanating from
the nominal interaction point.

\subsection{Measurement of the scattered positron}

The combination of calorimetric and tracking information for the
measurement of the four-momentum of the scattered positron resulted in
significant improvements with respect to the previous analysis. The
BPT tracking information proved crucial in suppressing beam- and
neutral-particle-related backgrounds. It also allowed a better control
of the systematic uncertainties in the position-dependent corrections
to the BPC energy measurement, the development of an improved position
reconstruction algorithm for the BPC, a direct determination of the
fiducial area of the measurement, and a significant reduction of the
uncertainty in the position of the BPC\@. Moreover, the fact that the
longitudinal coordinate of the event vertex was reconstructed from the
scattered-positron track reduced the dependence of the reconstruction
of kinematic variables on the hadronic final state.

The energy deposited in the BPC by the scattered positron was
reconstructed as the sum over all energies measured in the BPC
fingers. Calibration constants for individual stacks of fingers were
determined using kinematic-peak events\footnote{A cut
$y_\mathrm{JB}<0.04$ (see Section~\ref{ss:hfs}) selects events for
which the energy of the scattered positron sharply peaks within 2\% of
the beam energy, providing a good calibration source (kinematic-peak
events).}, taking into account position-dependent corrections for
transverse shower leakage from the BPC, light attenuation in the
scintillator fingers, and the non-uniformity caused by the gaps
between fingers \cite{CATHESIS}. After the calibration procedure, the
BPC energy response was found to be uniform to within $0.3\%$, and the
absolute energy scale at $27.5\GeV$ was known to $0.3\%$. The
linearity of the energy response was estimated through a simulation
based on the results of a scan with a $^{60}$Co source; it was found
that the energy scale at $4\GeV$ was accurate to within $1\%$
\cite{UFIBTHESIS}. The results of the simulation were confirmed by a
study of QED Compton scattering events \cite{VMTHESIS}.

The transverse position of a positron shower in the BPC was
reconstructed by using the correlation between the shower position
relative to the center of the struck stack of BPC fingers and the
energy deposited in its two neighbors. The function correlating the
distribution of energy to the shower position was determined by
comparing positron positions reconstructed separately with the BPC and
the BPT, respectively \cite{CATHESIS}. The resulting average BPC
position resolution was $\Delta X=\Delta Y=0.22\cm/\sqrt{\EGeV}$.

The projection of the positron scattering angle on the $X$-$Z$ plane,
$\theta_X$, the positron impact point on the BPC front face in $X$,
and the longitudinal position of the event vertex, $Z_\mathrm{vtx}$,
were determined by reconstructing a BPT track as the straight line
joining the hits in the two planes and assuming the average value for
the $X$ coordinate of the event vertex, determined with the Central
Tracking Detector (CTD, see below). A study of simulated events showed
that the effect of the magnetic field was negligible. The value of
$\theta_Y$ could only be reconstructed from the shower position in the
BPC\@. However, since for positrons scattered into the fiducial area
$\theta_X$ was significantly larger than $\theta_Y$, the impact of the
limited $\theta_Y$ resolution on the reconstruction of
$\theta_e=\sqrt{\theta_X^2+\theta_Y^2}$ was small. On average, the
angular resolution was $\Delta\theta_e=0.2\mrad$, and the vertex
resolution was $\Delta Z_\mathrm{vtx}=3\cm$.

The $X$ positions of the BPC and BPT were determined
with a minimization procedure based on the comparison between
the event vertex reconstructed from the BPC and BPT information
and that reconstructed from the CTD tracking information \cite{CATHESIS}.
Several studies were performed to determine the stability of
the result: the largest contribution to the systematic error comes from the
uncertainty in the relative $Z$ position of the BPT planes, estimated from a
survey to be $300\mum$.
The resulting total accuracy in the $X$ positions of the BPC and BPT is
$200\mum$.
The tracking efficiency of the BPT was $90.4\pm 1.5\%$, the inefficiency
being primarily due to dead strips.

\subsection{Measurement of the hadronic final state} \label{ss:hfs}

The hadronic final state was measured using the uranium-scintillator
calorimeter (CAL) and the CTD\@. The CAL covers 99.7\% of the total
solid angle. Under test-beam conditions, its relative energy
resolution is ${\Delta E}/{E}={18\%}/{\sqrt{\EGeV}}$ for
electromagnetic showers and ${\Delta E}/{E}={35\%}/{\sqrt{\EGeV}}$ for
hadronic showers. The CTD operates in a $1.43\,\mathrm{T}$ solenoidal
magnetic field, and has a relative resolution for full-length tracks
of ${\Delta p_T}/{p_T}=0.0058\cdot p_T\, \oplus 0.0065\, \oplus
0.0014/p_T$, with $p_T$ measured in Ge\kern-1pt V (the symbol $\oplus$
denotes addition in quadrature). The interaction vertex is measured
with a typical resolution along (transverse to) the beam direction of
$4\mm$ ($1\mm$).

Global variables characterizing the hadronic final state were measured
by summing all reconstructed hadronic final-state objects. A
combination of clusters of energy deposits in the CAL and the
corresponding tracks measured in the CTD \cite{ZUFO} was found to
provide the best resolution in determining the quantities
\begin{equation}
\begin{array}{c} \vspace*{\belowdisplayskip}\displaystyle
p_{X}^{\mathrm{had}} = \sum_iE_i\sin\vartheta_i\cos\phi_i , \quad\quad
p_{Y}^{\mathrm{had}} = \sum_iE_i\sin\vartheta_i\sin\phi_i , \quad\quad
p_{Z}^{\mathrm{had}} = \sum_iE_i\cos\vartheta_i , \\ \displaystyle
E^{\mathrm{had}} = \sum_i E_i , \quad\quad p_{T}^{\mathrm{had}} =
\sqrt{\left(p_{X}^{\mathrm{had}}\right)^2+\left(p_{Y}^{\mathrm{had}}\right)^2} ,
\quad\quad \delta^{\mathrm{had}} =
E^{\mathrm{had}}-p_{Z}^{\mathrm{had}} , \quad\quad y_\mathrm{JB} =
\frac{\delta^{\mathrm{had}}}{2E_e} .
\label{e:HFS} \end{array}
\end{equation}

\subsection{Kinematic reconstruction} \label{ss:kin_rec}

Two methods have been used for the reconstruction of the kinematic
variables $Q^2$ and $y$; the variable $x$ was then determined from
$x=Q^2/(sy)$. The first method uses the energy, $E_e'$, and angle,
$\theta_e$, of the scattered positron (``electron method'') to
determine
\begin{equation}
\begin{array}{ccccl} \vspace*{\belowdisplayskip} Q^2_e & = &
2E_eE_e'(1+\cos\vartheta_e) & \approx & E_eE_e'\theta_e^2, \\ y_e & =
& 1 - {\displaystyle \frac{E_e'}{2E_e}} (1-\cos \vartheta_e) & \approx
& 1 - {\displaystyle \frac{E_e'}{E_e}}. \end{array}
\label{e:ELMETHOD}
\end{equation}

The second method is used for low values of $y$, where the $y$
resolution of the electron method can be improved by combining the
scattered-positron variables with hadronic final-state variables
(``$e\Sigma$ method'' \cite{BASBERN}) to determine
\begin{equation}
\begin{array}{ccl} \vspace*{\belowdisplayskip}
Q^2_{e\Sigma}&=& Q^2_e,\\ y_{e\Sigma}
&=&\displaystyle\frac{2E_e}{\delta}
\cdot\frac{\delta^{\mathrm{had}}}{\delta} \, ,
\end{array} \label{ESIMETHOD}
\end{equation}
where $\delta$ is defined as $\delta = \delta^{\mathrm{had}}
+E_e'(1-\cos\vartheta_e)$. Four-momentum conservation requires
$\delta=2E_e$ for fully contained and perfectly measured events.

In the analysis, the kinematic variables of the events were
reconstructed with the electron method for $y_e>0.08$, and with the
$e\Sigma$ method for $y_{e\Sigma}<0.08$. The resulting relative
resolution in $y$ ranged from about 5\% at high $y$ to 25\% at low
$y$. The relative $Q^2$ resolution was about 10\% in the entire
kinematic range of the present measurement. The use of the $e\Sigma$
method allowed the extension of the measured kinematic range to $y$
values as low as $0.005$.

\subsection{Luminosity measurement}

The luminosity was measured from the rate of photons from the
Bethe-Heitler process, $ep\to ep\gamma$, using a lead-scintillator
calorimeter \cite{LUMI} positioned at $Z = -107\m$ to detect photons
scattered through angles of less than $0.5\mrad$. The data set used in
the analysis corresponded to an integrated luminosity of $3.9\pb$,
with an uncertainty of $1.8\%$.

\section{Trigger, event selection and background}

The event selection was based mainly on the requirement of a
well-reconstructed positron in the BPC and in the BPT, while
additional cuts on the hadronic final state suppressed background and
limited effects of resolution smearing (event migrations) and
radiative corrections.

Events were selected online by the ZEUS three-level trigger system.
The trigger required a minimum energy deposit in the BPC, a timing
compatible with an $ep$ interaction, and imposed requirements on
energy deposits from the hadronic final state in the CAL\@. No BPT
information was used in the trigger.

The offline event selection imposed tighter requirements than those
applied in the trigger. Using an independently triggered control
sample, the trigger was found to be more than 99.5\% efficient for
events passing the offline cuts.

The scattered positron was required to have an energy, $E_e'$, of at
least $4.4\GeV$ for events reconstructed with the electron method, or
at least $20\GeV$ for the $e\Sigma$ method. The positron position at
the BPC front face as extrapolated from the BPT measurement had to lie
within the fiducial area. In order to identify electromagnetic showers
and reject hadrons, the transverse size (energy-weighted r.m.s.) of
the shower in the BPC was required to be less than $0.8\cm$. In order
to limit migrations from low $y$, a cut $y_\mathrm{JB}>0.06$ was
applied for events reconstructed with the electron method
($y_\mathrm{JB}>0.004$ for the $e\Sigma$ method).

Background came primarily from events in which a particle from the
hadronic final state, usually a photon from the decay of a $\pi^0$,
produced an electromagnetic shower in the BPC, while the scattered
positron escaped undetected. Such events originated mostly from
photoproduction (with $Q^2$ lower than measured in the present
analysis). In addition, photons from initial- and final-state
radiation (ISR and FSR) could produce an electromagnetic shower in the
BPC\@. Since the longitudinal momentum carried away by the undetected
particles (usually escaping through the rear beam-pipe hole) decreases
$\delta$, a cut $\delta>30\GeV$ was applied to reject both
photoproduction background and radiative events with a hard
initial-state photon. In addition, background from neutral particles
was suppressed through the requirement that the extrapolated track
from the BPT match the position of the BPC shower in $X$ within five
times the average BPC position resolution. The remaining contamination
was determined from the number of events in which the scattered
positron was detected in an electromagnetic calorimeter positioned at
$Z = -35\m$, designed to tag positrons with $Q^2<0.01\GeV^2$ scattered
through very low angles, scaled by the calorimeter tagging efficiency
(estimated using a sample of simulated photoproduction events passing
all selection cuts). The background contamination was found to be less
than 1.5\% at high $y$ and negligible at low $y$, and was
statistically subtracted from the data sample. The uncertainty
associated with this background correction was estimated by comparing
it to the number of simulated photoproduction events (in a
luminosity-normalized sample) which passed all analysis cuts and
taking the difference to be the uncertainty.

In order to suppress beam-related background, the longitudinal
coordinate of the event vertex as reconstructed by the BPT was
required to lie within $90\cm$ of the nominal interaction point. A cut
$\delta<65\GeV$ rejected events from processes other than $ep$
interactions. The fraction of such background events, determined from
events recorded at times when there was no $ep$ beam crossing, was
found to be below $0.5\%$ and was neglected.

\section{Analysis}

\subsection{Monte Carlo simulation}

Monte Carlo (MC) simulations were used to characterize the accuracy of
the kinematic reconstruction, to determine the efficiency of selecting
events, to estimate the background rate, and to extract $\ft$.
Non-diffractive processes including first-order QED radiative
corrections were simulated using the HERACLES~4.6.1 program with the
DJANGOH~1.1 interface \cite{HERACLES+HSDJANGOH} to the QCD
programs. The program RAPGAP~2.06 \cite{HJRAPGAP} was used to simulate
diffractive processes in which the incoming proton emits a Pomeron
that has partonic structure. The color-dipole model of ARIADNE~4.08
\cite{ARIADNE} was used for the simulation of the parton shower, while
the hadronization was based on the Lund string model of JETSET~7.410
\cite{JETSET}. Photoproduction background events were simulated using
PYTHIA~5.724 \cite{PYTHIA}. All generated events were passed through a
detector simulation based on GEANT~3.13 \cite{GEANT}.

In order to improve the simulation of the hadronic final state, the
diffractive and non-diffractive samples were mixed in a proportion
determined from the data (while maintaining a fixed overall MC
normalization). This was crucial at high $y$ and low $Q^2$, where the
trigger and offline event selection efficiencies for diffractive and
non-diffractive events were significantly different, owing to the
different topologies of the hadronic final state. The mixing ratio was
determined by optimizing the agreement between data and MC in
distributions of hadronic final-state quantities according to a
least-squares minimization procedure applied to histograms of these
quantities \cite{CATHESIS}. The hadronic variables chosen for the
determination of the mixing ratio, parameterized\footnote{Although in
principle the mixing ratio between the diffractive and non-diffractive
Monte Carlo samples may depend on both $x$ and $Q^2$, it was found
that it could be parameterized with sufficient accuracy as a function
of $x$ only.} as a function of $x$, were the pseudorapidity of the
most forward hadronic energy deposit or track, $\emax$, and the
variables $\delta^{\mathrm{had}}$ and $p_{T}^{\mathrm{had}}$ in
combination. The average of the two results for the mixing fraction
was used for the extraction of $\ft$, while the difference between
them was taken into account as a contribution to the systematic
uncertainties (see Section \ref{sec:SYS}). The fraction of diffractive
events used for the extraction of $\ft$ varied between 10\% and 25\%
over the $x$ range of the measurement.

\begin{figure}[ht]

{\centering \setlength{\unitlength}{1cm}\begin{picture}(12.8,20.33)
\put(0,0){\epsfig{figure=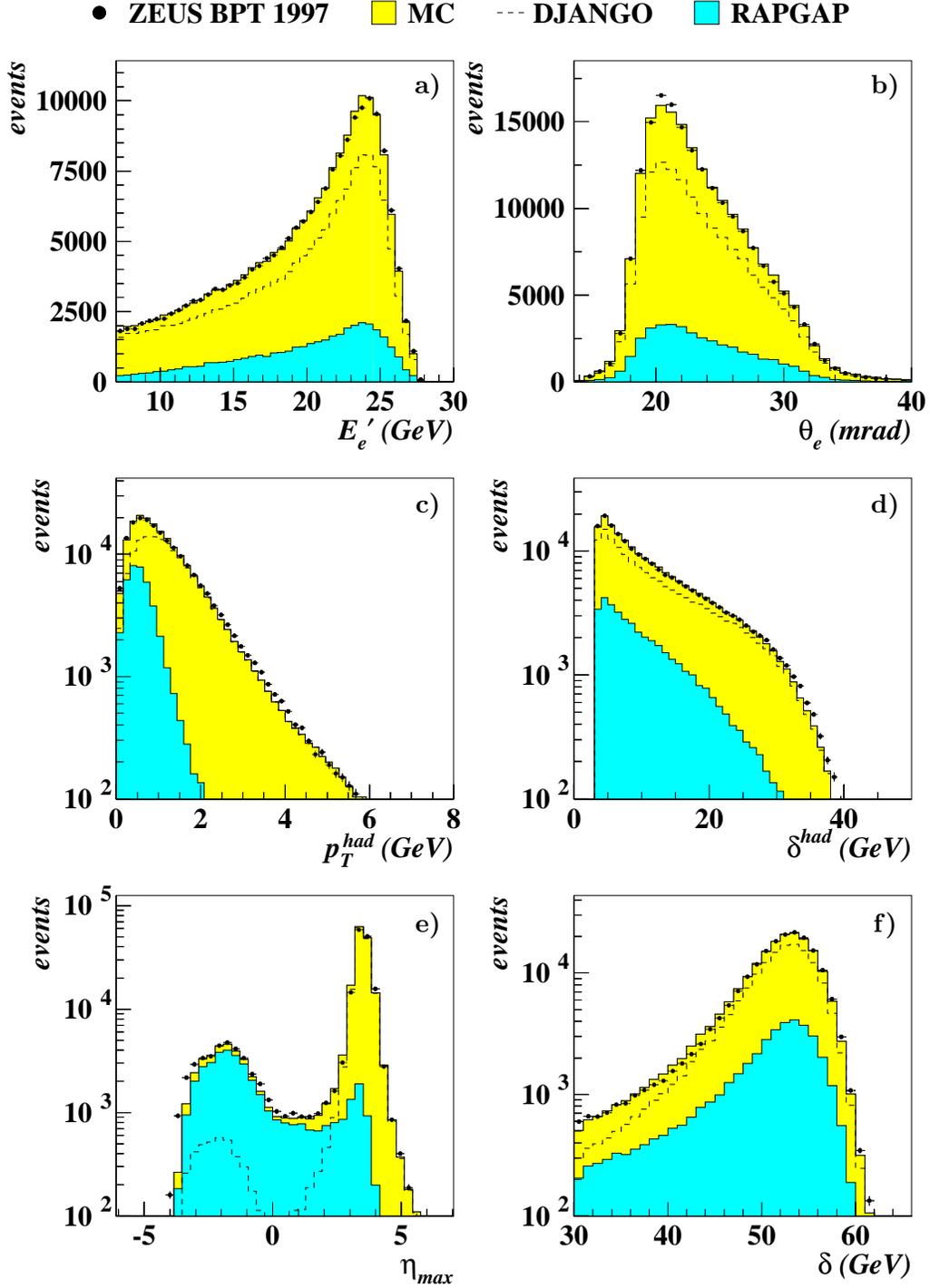,width=13.5cm}}
\SetScale{28.346457} \SetWidth{0.0176389}

\Text(6.2,17.55)[]{\bf a)} \Text(6.2,11.4)[]{\bf c)}
\Text(6.2,5.25)[]{\bf e)} \Text(12.9,17.55)[]{\bf b)}
\Text(12.9,11.4)[]{\bf d)} \Text(12.9,5.25)[]{\bf f)}

\end{picture}

\caption{\label{CONTROL1}Distributions of reconstructed quantities in
data and MC simulation in the region defined by $0.06<y_\mathrm{JB}$
and $y_e<0.74$: a) energy, $E_e'$, of the positron measured in the
BPC; b) scattering angle, $\theta_e$, of the positron measured in the
BPT; c) $p_T^\mathrm{had}$; d) $\delta^\mathrm{had}$; e) $\emax$; f)
$\delta$. The points denote measured data, the light shaded histogram
is the sum of non-diffractive (DJANGO) and diffractive (RAPGAP) MC,
the dark shaded histogram and the dashed line represent the individual
contributions from RAPGAP and DJANGO, respectively.}}

\end{figure}

\begin{figure}[ht]

{\centering \setlength{\unitlength}{1cm}\begin{picture}(13.5,14.81)
\put(0,0){\epsfig{figure=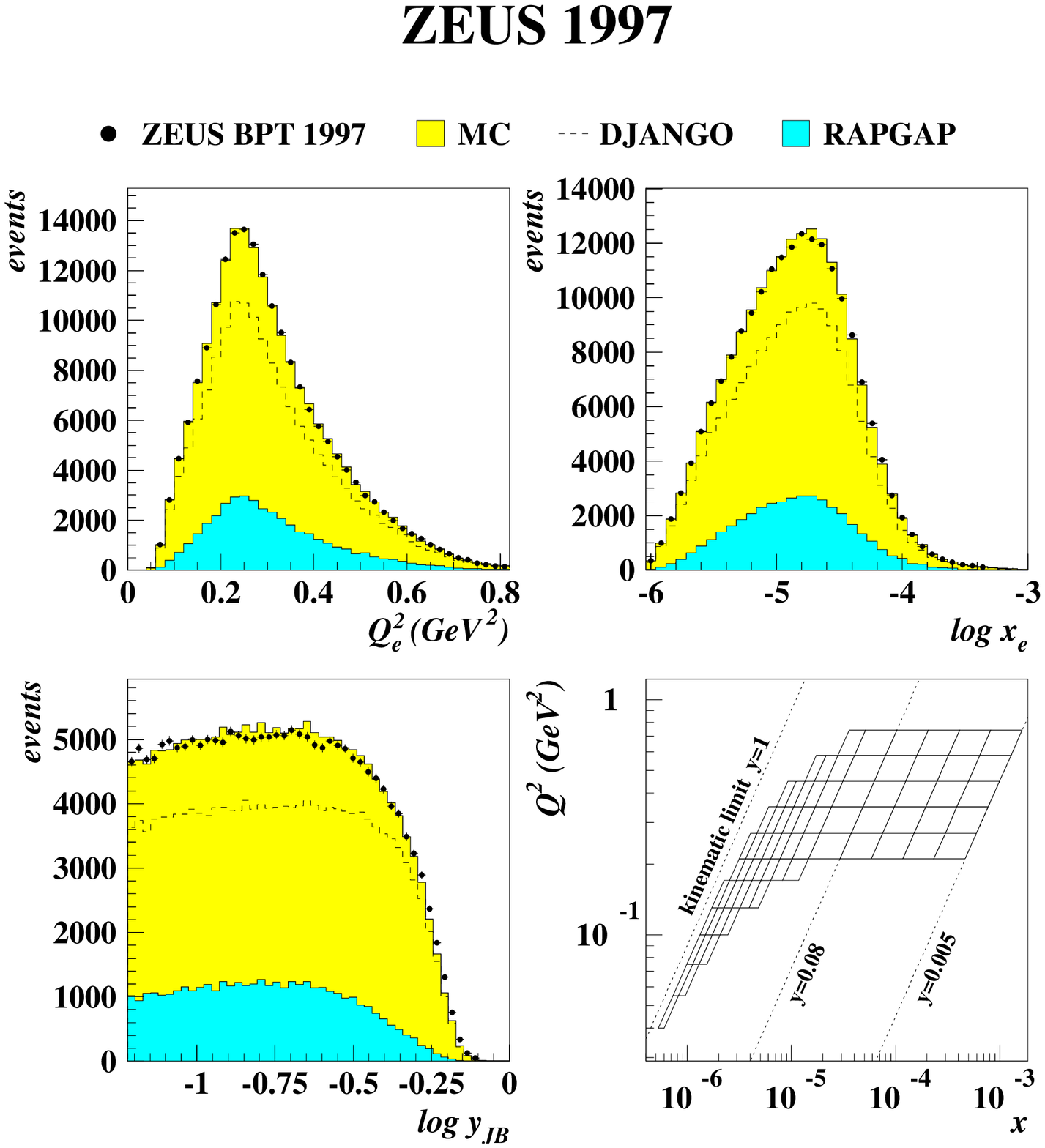,width=13.5cm}}
\SetScale{28.346457} \SetWidth{0.0176389}

\Text(6.2,12.05)[]{\bf a)} \Text(6.2,5.65)[]{\bf c)}
\Text(12.9,12.05)[]{\bf b)} \Text(12.9,5.65)[]{\bf d)}

\end{picture}

\caption{\label{CONTROL2}a)--c) Distributions of reconstructed
kinematic quantities in data and MC simulation in the region defined
by $0.06<y_\mathrm{JB}$ and $y_e<0.74$: a) $Q^2_e$; b) $x_e$; c)
$y_\mathrm{JB}$. The points denote measured data, the light shaded
histogram is the sum of non-diffractive (DJANGO) and diffractive
(RAPGAP) MC, the dark shaded histogram and the dashed line represent
the individual contributions from RAPGAP and DJANGO, respectively. d)
The bins in the kinematic plane ($Q^2$ vs.\ $x$). The bin boundaries
in $y$ are 0.005, 0.01, 0.02, 0.04, 0.08, 0.16, 0.23, 0.30, 0.37,
0.45, 0.54, 0.64, 0.74, 0.84; those in $Q^2$ are 0.040, 0.055, 0.075,
0.10, 0.13, 0.17, 0.21, 0.27, 0.35, 0.45, 0.58, $0.74\GeV^2$. The
border at $y=0.08$, above which the electron method and below which
the $e\Sigma$ method was used to reconstruct the event kinematics, is
also indicated.}}

\end{figure}

Figure \ref{CONTROL1} shows comparisons between distributions of
reconstructed positron and hadronic final-state variables in data and
simulation for $0.06<y_\mathrm{JB}$ and $y_e<0.74$. Figure
\ref{CONTROL2}a--c shows comparisons between distributions of
reconstructed kinematic variables. Good agreement between data and
simulation was obtained in this region and the agreement is of similar
quality in the remaining kinematic region used for the extraction of
$\ft$.

\subsection{Binning of the data}

Figure\ \ref{CONTROL2}d shows the bin boundary grid used to extract
$\ft$. The bin widths in $y$ are approximately three times the $y$
resolution at low $y$, and twice the resolution at higher $y$. The bin
widths in $Q^2$ are about four times the $Q^2$ resolution. In the
region of overlap, the bin boundaries were chosen to be identical to
those used in the previous measurement \cite{ZEUSBPC}.

The geometrical acceptance (defined as the fraction of events where
the scattered positron hits the fiducial area) was determined from the
Monte Carlo simulation to be between 4\% and 14\% for most of the
bins, decreasing to 1\% for a few bins. The acceptance (defined as the
fraction of events passing all selection cuts and generated in a bin
that were also reconstructed in that bin) varied between 50\% and
70\%, and the purity (defined as the fraction of events passing all
selection cuts and reconstructed in a bin that were also generated in
that bin) typically ranged from 40\% to 65\%, decreasing to 30\% at
the lowest $y$.

\subsection{\boldmath Extraction of $\ft$}

The proton structure function $\ft$ was measured using an iterative
bin-by-bin unfolding method. The measured value of $\ft$ at a given
point $(x,Q^2)$ inside a bin was obtained as the ratio of the numbers
of events reconstructed in the bin in data and MC, multiplied by the
corresponding $\ft$ value of the parameterization used in the MC
sample. The point $(x,Q^2)$ at which $\ft$ is quoted was chosen such
that $y$ and $Q^2$ are round numbers close to the center of the bin.

First-order QED radiative corrections were included in the MC
simulation used for the unfolding. The uncertainty in these
corrections is due principally to the uncertainty in $\ft$ at $x$ and
$Q^2$ values outside the region of this measurement, which was
estimated to be 6\% from an analysis of events with ISR
\cite{CATHESIS} and was taken into account in the systematic error
calculation. Higher-order QED radiative corrections, including
soft-photon exponentiation, were evaluated using the program HECTOR
\cite{HECTOR} in the leading-log approximation. They were found to be
less than 0.2\% and were thus neglected.

Since the bin-by-bin unfolding requires a precise simulation of
migration effects, which in turn requires $\ft$ to be approximately
the same in data and MC, the MC sample was iteratively re-weighted by
the extracted $\ft$. The contribution from $\fl$ was neglected in this
procedure. The MC events were weighted according to the ALLM97
\cite{ALLM97} parameterization in the first iteration, and according
to a function of the form\footnote{This phenomenological
parameterization is based on the combination of a simplified version
of the generalized vector meson dominance model \cite{GVDMTHEORY} for
the description of the $Q^2$ dependence and Regge theory
\cite{REGGETHEORY} for the description of the $x$ dependence of
$\ft$.} \cite{ZEUSSVX95}
\begin{equation}
\ft(x,Q^2) = \left(\frac{Q^2}{4\pi^2\alpha}\right)\cdot
\left(\frac{M_0^2}{M_0^2+Q^2}\right)\cdot \left(A_\reg\cdot
\left(\frac{Q^2}{x}\right)^{\alpha_\reg-1}+A_\pom\cdot
\left(\frac{Q^2}{x}\right)^{\alpha_\pom-1}\right)
\label{e:GVDM+REGGE}
\end{equation}
in further steps. At each iteration, $\alpha_\reg$ was fixed to the
value $0.5$ and the values of the remaining four parameters in Eq.\
(\ref{e:GVDM+REGGE}) were obtained from a fit to the results obtained
in the previous step; in order to constrain the fit at high $x$,
photoproduction data \cite{DOCALDWELL} were also included\footnote{For
this purpose, Eq.\ (\ref{e:GVDM+REGGE}) was re-written using
$\sgsp=(4\pi^2\alpha/Q^2)\ft$ and $W^2=Q^2/x$, where $W$ denotes the
photon-proton center-of-mass energy.}. The $\ft$ results converged
after a few iterations.

To correct for the effect of $\fl$, the value of $R = \fl/(\ft-\fl)$
from the BKS model \cite{BKSMODEL} was used, which, in the range of
interest here, can be parameterized to a very good approximation by
$R=0.165\cdot Q^2/m_\rho^2$, where $m_\rho=0.77\GeV$ is the $\rho$
meson mass. This changed the extracted $\ft$ in the bins of the
present measurement by at most 3\% with respect to the values
determined assuming $\fl=0$.

\subsection{Systematic uncertainties} \label{sec:SYS}

The systematic uncertainties in the measured $\ft$ values were
determined by studying the stability of the results under variations
of the reconstruction parameters entering detector calibration and
alignment, of the simulated detector efficiency, of the event
selection cuts, and of several other aspects of the analysis
procedure. The reconstruction parameters were varied within their
uncertainties. The variations of the selection cuts took into account
the degree of arbitrariness in the cut values. The following checks
were performed (the effect on $\ft$ is given in parentheses):

\begin{itemize}

\item Analysis cuts:

\begin{itemize}

\item variation of the cut $\delta>30\GeV$ by $\pm\, 2\GeV$ (mostly
below 1.5\%, up to 4\% at high $y$);

\item variation of the cut $y_\mathrm{JB}>0.06$ by $\pm 0.01$, and of
the cut $y_\mathrm{JB}>0.004$ by $\pm 0.001$ (up to 1.5\% at medium
$y$);

\item variation of the BPC shower-width cut at $0.8\cm$ by $\pm
0.1\cm$ (up to 1.5\% at low positron energies);

\item variation of the $5\sigma$ BPC shower/BPT track-match cut by
$\pm 2\sigma$ (up to 1.5\% at low positron energies);

\item change of the BPT vertex cut from $\pm 90\cm$ to $\pm 50\cm$ (up
to 3.5\% in bins with low geometrical acceptance);

\item variation of the fiducial cut in $X$ by $\pm 1\mm$ (mostly below
1.5\%, up to 6\% in bins with low geometrical acceptance);

\item variation of the fiducial cut in $Y$ by $\pm 1\mm$ (up to
2.5\%).

\end{itemize}

\item Detector calibration, alignment, and efficiency:

\begin{itemize}

\item variation of the BPC energy scale by $\pm 0.3\%$ (up to 2.5\% at
low $y$);

\item variation of the BPC energy response by $\pm 1\%$ at $4\GeV$,
decreasing linearly to $0\%$ at $27.5\GeV$ (up to 2.5\% at medium and
high $y$);

\item variation of the absolute BPC/BPT position in $X$ by $\pm
200\mum$ (mostly 1.5\%, up to 3.5\% in some bins);

\item variation of the simulated BPT efficiency by $\pm 1.5\%$
(1.5\%);

\item variation of the CAL energy scale for hadrons by $\pm 3\%$ (up
to 2.5\% at high and low $y$).

\end{itemize}

\item Analysis procedure:

\begin{itemize}

\item variation of the fractions of DJANGO and RAPGAP events in the MC
sample by half of the difference between fractions obtained from
separate optimizations in $\emax$ or in $\delta^{\mathrm{had}}$ and
$p_{T}^{\mathrm{had}}$ (mostly below 1\%, up to 8\% in the two highest
$y$ bins);

\item variation of the subtracted photoproduction background by
$+200\%$ and $-100\%$ (up to 2.5\% at high $y$);

\item re-weighting of $\ft$ in the MC sample, outside the bins used
for the measurement, by $\pm 6\%$, which induces a variation of the
number of events migrating into the measurement region\footnote{This
check is particularly sensitive to the magnitude of radiative
corrections due to migrations of unrecognized ISR events from higher
$x$ and lower $Q^2$, but also to the amount of migrations due to
detector resolution effects.} (up to 3\% in some bins).

\end{itemize}

\end{itemize}

The total systematic error was computed as the quadratic sum of the
individual contributions, separately for positive and negative
deviations. The average statistical error is $2.6\%$ and the average
systematic error $3.3\%$. In most bins, the systematic error has a
magnitude similar to that of the statistical error. No individual
contribution to the systematic error dominates, except in the two
highest $y$ bins, where the error is dominated by the uncertainty in
the fraction of diffractive events. An additional overall
normalization uncertainty of $1.8\%$ due to the luminosity measurement
is not included in the systematic error.

\section{Results}

\begin{figure}[hp]

{\centering

\epsfig{figure=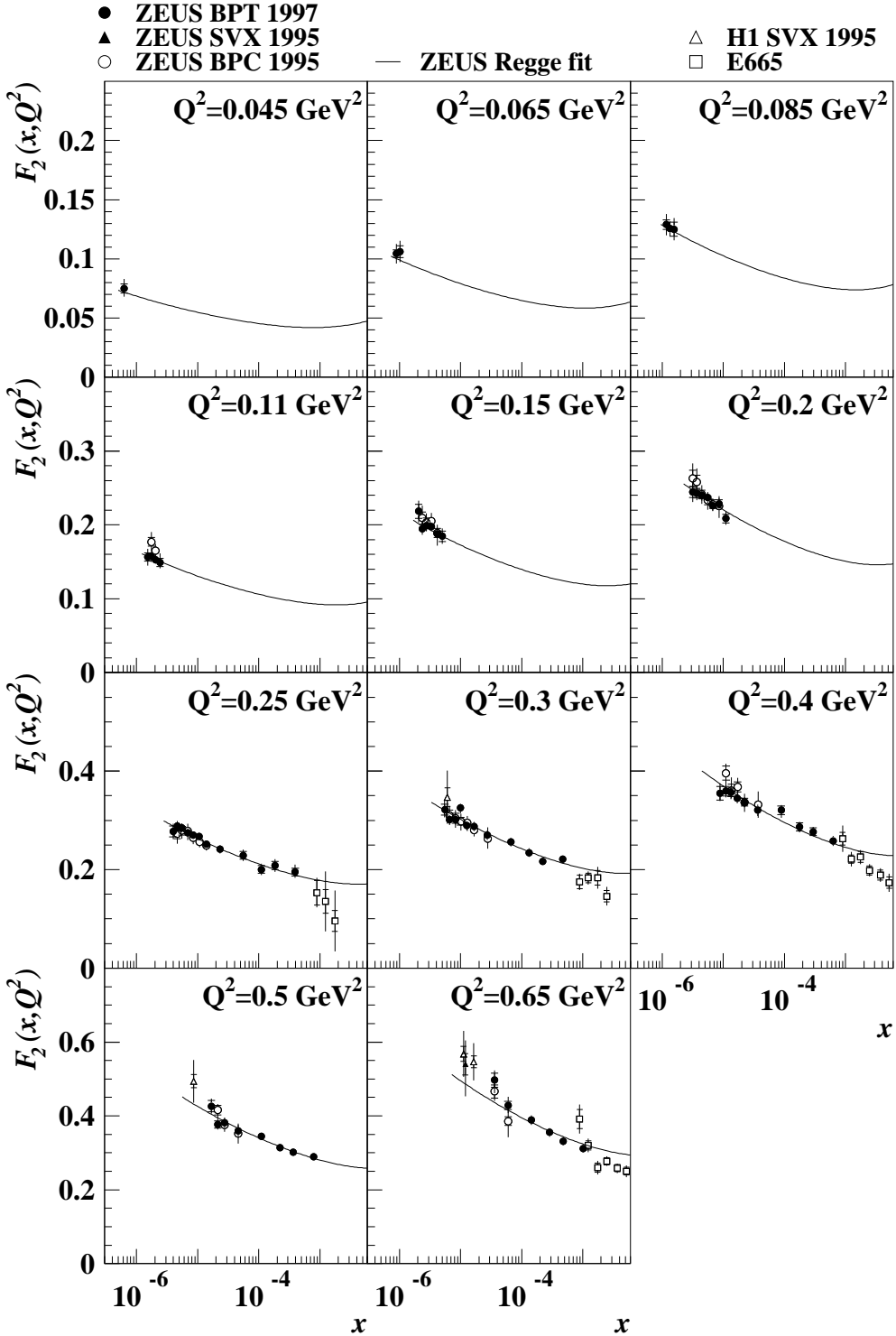,width=13.1cm}

\caption{\label{FTWOBIGA}Measured $\ft$ vs.\ $x$ in bins of $Q^2$. The
data from the present measurement are indicated by filled circles. The
solid line shows the ZEUS Regge fit. Open circles denote the results
from a previous analysis, filled and open triangles denote other
measurements from ZEUS and H1, respectively, and squares denote
results from E665. These other measurements have been shifted to the
$Q^2$ values of the present measurement using the ALLM97
parameterization. The inner error bars represent statistical errors,
the outer ones the sum in quadrature of statistical and systematic
errors; normalization uncertainties are not included.}}

\end{figure}

The measurement of $\ft$ presented here uses data in the kinematic
region $0.04\GeV^2<Q^2<0.74\GeV^2$ and $5.3\cdot 10^{-7}<x<1.6\cdot
10^{-3}$, corresponding to $0.005<y<0.84$. The values of $\ft$
extracted in 70 bins are listed in Table \ref{FTWOVALUES}. Figure\
\ref{FTWOBIGA} shows these $\ft$ values as a function of $x$ for
different bins of $Q^2$, together with previous ZEUS and H1 data at
low $Q^2$ \cite{ZEUSBPC,ZEUSSVX95,H1SVX95} and with data at higher $x$
from the fixed-target experiment E665 \cite{E665}. The curve denoted
as ``ZEUS Regge fit'' represents the parameterization of Eq.\
(\ref{e:GVDM+REGGE}), with the values of the parameters resulting from
the last iteration in the extraction of $\ft$: $A_\reg=147.8\pm
4.6\mub$, $\alpha_\reg= 0.5$, $A_\pom=62.0\pm 2.3\mub$,
$\alpha_\pom=1.102\pm 0.007$, $M_0^2=0.52\pm 0.04\GeV^2$ (statistical
and systematic errors added in quadrature).
These results are in good agreement with those previously
obtained \cite{ZEUSSVX95}.

The errors in $\ft$ are significantly reduced compared to the previous
measurement \cite{ZEUSBPC}. While in general good agreement is
observed in the region of overlap, the present results are slightly
lower than the previous ones at the lowest $Q^2$ and low $x$. At the
highest $x$ of this measurement, the present data overlap with data
from E665. Although the $\ft$ values do not differ significantly
in the region of overlap, if the ZEUS Regge fit is extrapolated to
higher $x$ into the E665 region, it is found to lie about 15\%
above the E665 values.

The rise of the proton structure function $\ft$ at low $x$, observed
to be steep in HERA data at higher $Q^2$, persists down to small
$Q^2$, but becomes shallower as $Q^2$ decreases into the range of the
present measurement. At the low $Q^2$ values of this measurement, the
rise of $\ft$ at low $x$ is well described by Regge theory assuming a
constant logarithmic slope $\partial{\ln}F_2/\partial{\ln}(1/x)$,
as reflected in the good agreement between the data and the ZEUS Regge fit.

\begin{figure}[hp]

{\centering
\epsfig{figure=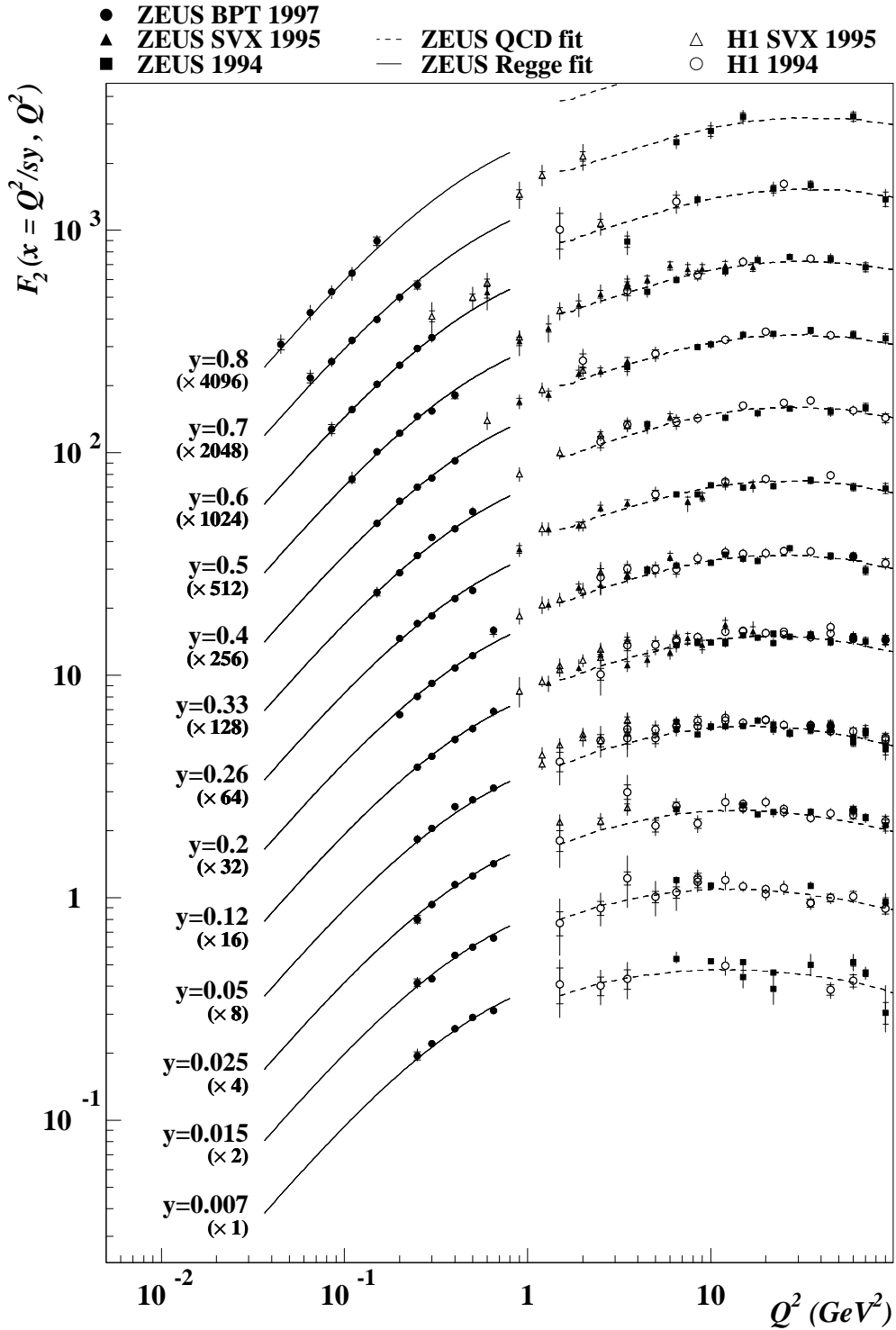,width=13.1cm}

\caption{\label{SIGTOTQ}Measured $\ft$ vs.\ $Q^2$ in bins of $y$. The
data from the present measurement are indicated by filled
circles. Triangles, filled squares and open circles denote other
measurements from ZEUS and H1. The data have been scaled by the
numbers in parentheses for clarity of presentation. The solid line at
low $Q^2$ shows the ZEUS Regge fit, the dashed line at higher $Q^2$
the ZEUS QCD fit. The other measurements have been shifted to the $y$
values of the present measurement using the ALLM97 parameterization.}}

\end{figure}

Figure\ \ref{SIGTOTQ} shows $\ft$ as a function of $Q^2$ for different
bins of $y$, together with previous ZEUS and H1 data
\cite{ZEUS+H194,ZEUSSVX95,H1SVX95}. At higher $Q^2$, $\ft$ is roughly
independent of $Q^2$. The curve denoted as ``ZEUS QCD fit''
\cite{ZEUSSVX95} illustrates that this behavior is well described by
next-to-leading-order QCD fits down to $Q^2$ values of about
$1\GeV^2$. On the other hand, it is clear that $\ft$ must acquire a
stronger $Q^2$-dependence at sufficiently low $Q^2$, since
conservation of the electromagnetic current requires $\ft$ to vanish
like $Q^2$ as $Q^2\to 0$. Dynamical mechanisms, e.g.\ parton
saturation at small $x$ values \cite{SATURATION}, can
produce such a behavior at low $Q^2$. The present data exhibit a
smooth transition to a stronger $Q^2$ dependence in the $Q^2$ range
between $0.1\GeV^2$ and $1\GeV^2$, approaching, at the lowest $Q^2$
values of this measurement, a region where $\ft$ becomes nearly
proportional to $Q^2$. This transition is well described by the ZEUS
Regge fit.

\section{Summary}

The proton structure function $\ftxq$ has been measured in the
kinematic range $0.045\GeV^2<Q^2<0.65\GeV^2$ and $6\cdot
10^{-7}<x<1\cdot 10^{-3}$, using an $e^+p$ data sample corresponding
to an integrated luminosity of $3.9\pb$. The addition of a Beam Pipe
Tracker in front of the Beam Pipe Calorimeter and an enhanced
simulation of the hadronic final state have resulted in coverage of a
larger kinematic region and in improved statistical precision and
systematic accuracy compared to previous results obtained with the
Beam Pipe Calorimeter alone.

At the low $Q^2$ values of the present measurement, the proton
structure function $\ft$ rises more slowly with $x$ than observed in
HERA data at higher $Q^2$. This slow rise can be described by Regge
theory with a constant logarithmic slope $\partial{\ln}F_2/\partial{\ln}(1/x)$.
Furthermore, the $\ft$ data
presented here exhibit a stronger $Q^2$ dependence than observed at
higher $Q^2$, approaching, at the lowest $Q^2$ values of this
measurement, a region where $\ft$ becomes nearly proportional to
$Q^2$.

\section*{Acknowledgments}

We thank the DESY directorate for their strong support and
encouragement, and the HERA machine group for their diligent efforts.
We are grateful for the support of the DESY computing and network
services. The design, construction and installation of the ZEUS
detector have been made possible by the ingenuity and effort of many
people from DESY and home institutes who are not listed as authors. It
is a pleasure to thank H.~Spiesberger and H.~Jung for useful
discussions.

\begin{flushleft}

\end{flushleft}


\begin{table}[p] {\centering

\begin{sideways}

\begin{minipage}{10cm} {\footnotesize \begin{tabular}{ccccccc} \hline
$Q^2$ & $x$ & $y$ & $\ft$ & $\pm\,\mathrm{stat}$ & $+\,\mathrm{sys}$ &
$-\,\mathrm{sys}$ \\ $(\mathrm{GeV}^2)$ & & & & & & \\ \hline \\

$ 0.045 $&$ 6.21\cdot10^{-7} $&$ 0.800 $&$ 0.0749 $&$ 0.0039 $&$ 0.0072 $&$ 0.0059 $\\
$ 0.065 $&$ 1.02\cdot10^{-6} $&$ 0.700 $&$ 0.1060 $&$ 0.0051 $&$ 0.0076 $&$ 0.0071 $\\
$ 0.065 $&$ 8.97\cdot10^{-7} $&$ 0.800 $&$ 0.1043 $&$ 0.0032 $&$ 0.0077 $&$ 0.0077 $\\
$ 0.085 $&$ 1.56\cdot10^{-6} $&$ 0.600 $&$ 0.1250 $&$ 0.0060 $&$ 0.0075 $&$ 0.0070 $\\
$ 0.085 $&$ 1.34\cdot10^{-6} $&$ 0.700 $&$ 0.1255 $&$ 0.0029 $&$ 0.0055 $&$ 0.0061 $\\
$ 0.085 $&$ 1.17\cdot10^{-6} $&$ 0.800 $&$ 0.1289 $&$ 0.0039 $&$ 0.0083 $&$ 0.0081 $\\
$ 0.110 $&$ 2.43\cdot10^{-6} $&$ 0.500 $&$ 0.1489 $&$ 0.0056 $&$ 0.0108 $&$ 0.0053 $\\
$ 0.110 $&$ 2.02\cdot10^{-6} $&$ 0.600 $&$ 0.1530 $&$ 0.0033 $&$ 0.0053 $&$ 0.0047 $\\
$ 0.110 $&$ 1.73\cdot10^{-6} $&$ 0.700 $&$ 0.1569 $&$ 0.0033 $&$ 0.0059 $&$ 0.0065 $\\
$ 0.110 $&$ 1.51\cdot10^{-6} $&$ 0.800 $&$ 0.1565 $&$ 0.0056 $&$ 0.0095 $&$ 0.0102 $\\
$ 0.150 $&$ 5.02\cdot10^{-6} $&$ 0.330 $&$ 0.1844 $&$ 0.0070 $&$ 0.0078 $&$ 0.0074 $\\
$ 0.150 $&$ 4.14\cdot10^{-6} $&$ 0.400 $&$ 0.1880 $&$ 0.0046 $&$ 0.0057 $&$ 0.0041 $\\
$ 0.150 $&$ 3.31\cdot10^{-6} $&$ 0.500 $&$ 0.1976 $&$ 0.0037 $&$ 0.0044 $&$ 0.0051 $\\
$ 0.150 $&$ 2.76\cdot10^{-6} $&$ 0.600 $&$ 0.1986 $&$ 0.0037 $&$ 0.0052 $&$ 0.0058 $\\
$ 0.150 $&$ 2.36\cdot10^{-6} $&$ 0.700 $&$ 0.1947 $&$ 0.0045 $&$ 0.0068 $&$ 0.0069 $\\
$ 0.150 $&$ 2.07\cdot10^{-6} $&$ 0.800 $&$ 0.2183 $&$ 0.0097 $&$ 0.0106 $&$ 0.0118 $\\
$ 0.200 $&$ 1.10\cdot10^{-5} $&$ 0.200 $&$ 0.2085 $&$ 0.0058 $&$ 0.0072 $&$ 0.0061 $\\
$ 0.200 $&$ 8.49\cdot10^{-6} $&$ 0.260 $&$ 0.2286 $&$ 0.0051 $&$ 0.0057 $&$ 0.0045 $\\
$ 0.200 $&$ 6.69\cdot10^{-6} $&$ 0.330 $&$ 0.2260 $&$ 0.0046 $&$ 0.0047 $&$ 0.0050 $\\
$ 0.200 $&$ 5.52\cdot10^{-6} $&$ 0.400 $&$ 0.2372 $&$ 0.0046 $&$ 0.0057 $&$ 0.0057 $\\
$ 0.200 $&$ 4.41\cdot10^{-6} $&$ 0.500 $&$ 0.2399 $&$ 0.0049 $&$ 0.0056 $&$ 0.0055 $\\
$ 0.200 $&$ 3.68\cdot10^{-6} $&$ 0.600 $&$ 0.2423 $&$ 0.0057 $&$ 0.0067 $&$ 0.0066 $\\
$ 0.200 $&$ 3.15\cdot10^{-6} $&$ 0.700 $&$ 0.2442 $&$ 0.0074 $&$ 0.0066 $&$ 0.0109 $\\
$ 0.250 $&$ 3.94\cdot10^{-4} $&$ 0.007 $&$ 0.1950 $&$ 0.0081 $&$ 0.0125 $&$ 0.0078 $\\
$ 0.250 $&$ 1.84\cdot10^{-4} $&$ 0.015 $&$ 0.2082 $&$ 0.0083 $&$ 0.0098 $&$ 0.0094 $\\
$ 0.250 $&$ 1.10\cdot10^{-4} $&$ 0.025 $&$ 0.2004 $&$ 0.0079 $&$ 0.0086 $&$ 0.0079 $\\
$ 0.250 $&$ 5.52\cdot10^{-5} $&$ 0.050 $&$ 0.2289 $&$ 0.0079 $&$ 0.0083 $&$ 0.0080 $\\
$ 0.250 $&$ 2.30\cdot10^{-5} $&$ 0.120 $&$ 0.2411 $&$ 0.0034 $&$ 0.0080 $&$ 0.0076 $\\
$ 0.250 $&$ 1.38\cdot10^{-5} $&$ 0.200 $&$ 0.2513 $&$ 0.0037 $&$ 0.0056 $&$ 0.0062 $\\
$ 0.250 $&$ 1.06\cdot10^{-5} $&$ 0.260 $&$ 0.2676 $&$ 0.0042 $&$ 0.0060 $&$ 0.0064 $\\
$ 0.250 $&$ 8.36\cdot10^{-6} $&$ 0.330 $&$ 0.2698 $&$ 0.0048 $&$ 0.0065 $&$ 0.0077 $\\
$ 0.250 $&$ 6.90\cdot10^{-6} $&$ 0.400 $&$ 0.2744 $&$ 0.0052 $&$ 0.0054 $&$ 0.0065 $\\
$ 0.250 $&$ 5.52\cdot10^{-6} $&$ 0.500 $&$ 0.2848 $&$ 0.0059 $&$ 0.0064 $&$ 0.0073 $\\
$ 0.250 $&$ 4.60\cdot10^{-6} $&$ 0.600 $&$ 0.2883 $&$ 0.0071 $&$ 0.0072 $&$ 0.0093 $\\
$ 0.250 $&$ 3.94\cdot10^{-6} $&$ 0.700 $&$ 0.2776 $&$ 0.0114 $&$ 0.0105 $&$ 0.0110 $\\
\hline

\end{tabular}} \end{minipage}\hspace{0.5cm}\begin{minipage}{10cm}
{\footnotesize \begin{tabular}{ccccccc} \hline $Q^2$ & $x$ & $y$ &
$\ft$ & $\pm\,\mathrm{stat}$ & $+\,\mathrm{sys}$ & $-\,\mathrm{sys}$
\\ $(\mathrm{GeV}^2)$ & & & & & & \\ \hline \\

$ 0.300 $&$ 4.73\cdot10^{-4} $&$ 0.007 $&$ 0.2214 $&$ 0.0058 $&$ 0.0047 $&$ 0.0061 $\\
$ 0.300 $&$ 2.20\cdot10^{-4} $&$ 0.015 $&$ 0.2169 $&$ 0.0058 $&$ 0.0056 $&$ 0.0056 $\\
$ 0.300 $&$ 1.32\cdot10^{-4} $&$ 0.025 $&$ 0.2339 $&$ 0.0061 $&$ 0.0067 $&$ 0.0045 $\\
$ 0.300 $&$ 6.62\cdot10^{-5} $&$ 0.050 $&$ 0.2565 $&$ 0.0067 $&$ 0.0052 $&$ 0.0070 $\\
$ 0.300 $&$ 2.76\cdot10^{-5} $&$ 0.120 $&$ 0.2700 $&$ 0.0034 $&$ 0.0091 $&$ 0.0091 $\\
$ 0.300 $&$ 1.65\cdot10^{-5} $&$ 0.200 $&$ 0.2874 $&$ 0.0042 $&$ 0.0082 $&$ 0.0078 $\\
$ 0.300 $&$ 1.27\cdot10^{-5} $&$ 0.260 $&$ 0.2896 $&$ 0.0049 $&$ 0.0063 $&$ 0.0077 $\\
$ 0.300 $&$ 1.00\cdot10^{-5} $&$ 0.330 $&$ 0.3253 $&$ 0.0063 $&$ 0.0062 $&$ 0.0089 $\\
$ 0.300 $&$ 8.28\cdot10^{-6} $&$ 0.400 $&$ 0.3012 $&$ 0.0064 $&$ 0.0081 $&$ 0.0069 $\\
$ 0.300 $&$ 6.62\cdot10^{-6} $&$ 0.500 $&$ 0.3014 $&$ 0.0075 $&$ 0.0058 $&$ 0.0077 $\\
$ 0.300 $&$ 5.52\cdot10^{-6} $&$ 0.600 $&$ 0.3221 $&$ 0.0110 $&$ 0.0149 $&$ 0.0118 $\\
$ 0.400 $&$ 6.31\cdot10^{-4} $&$ 0.007 $&$ 0.2585 $&$ 0.0071 $&$ 0.0063 $&$ 0.0075 $\\
$ 0.400 $&$ 2.94\cdot10^{-4} $&$ 0.015 $&$ 0.2769 $&$ 0.0076 $&$ 0.0067 $&$ 0.0061 $\\
$ 0.400 $&$ 1.76\cdot10^{-4} $&$ 0.025 $&$ 0.2869 $&$ 0.0080 $&$ 0.0066 $&$ 0.0082 $\\
$ 0.400 $&$ 8.83\cdot10^{-5} $&$ 0.050 $&$ 0.3206 $&$ 0.0087 $&$ 0.0055 $&$ 0.0128 $\\
$ 0.400 $&$ 3.68\cdot10^{-5} $&$ 0.120 $&$ 0.3213 $&$ 0.0046 $&$ 0.0098 $&$ 0.0117 $\\
$ 0.400 $&$ 2.20\cdot10^{-5} $&$ 0.200 $&$ 0.3369 $&$ 0.0058 $&$ 0.0080 $&$ 0.0085 $\\
$ 0.400 $&$ 1.70\cdot10^{-5} $&$ 0.260 $&$ 0.3452 $&$ 0.0070 $&$ 0.0086 $&$ 0.0072 $\\
$ 0.400 $&$ 1.33\cdot10^{-5} $&$ 0.330 $&$ 0.3567 $&$ 0.0085 $&$ 0.0095 $&$ 0.0105 $\\
$ 0.400 $&$ 1.10\cdot10^{-5} $&$ 0.400 $&$ 0.3594 $&$ 0.0098 $&$ 0.0118 $&$ 0.0083 $\\
$ 0.400 $&$ 8.83\cdot10^{-6} $&$ 0.500 $&$ 0.3552 $&$ 0.0137 $&$ 0.0147 $&$ 0.0077 $\\
$ 0.500 $&$ 7.89\cdot10^{-4} $&$ 0.007 $&$ 0.2897 $&$ 0.0063 $&$ 0.0058 $&$ 0.0064 $\\
$ 0.500 $&$ 3.68\cdot10^{-4} $&$ 0.015 $&$ 0.3020 $&$ 0.0061 $&$ 0.0056 $&$ 0.0064 $\\
$ 0.500 $&$ 2.20\cdot10^{-4} $&$ 0.025 $&$ 0.3144 $&$ 0.0060 $&$ 0.0067 $&$ 0.0076 $\\
$ 0.500 $&$ 1.10\cdot10^{-4} $&$ 0.050 $&$ 0.3449 $&$ 0.0061 $&$ 0.0064 $&$ 0.0079 $\\
$ 0.500 $&$ 4.60\cdot10^{-5} $&$ 0.120 $&$ 0.3599 $&$ 0.0063 $&$ 0.0138 $&$ 0.0132 $\\
$ 0.500 $&$ 2.76\cdot10^{-5} $&$ 0.200 $&$ 0.3822 $&$ 0.0082 $&$ 0.0101 $&$ 0.0118 $\\
$ 0.500 $&$ 2.12\cdot10^{-5} $&$ 0.260 $&$ 0.3767 $&$ 0.0101 $&$ 0.0103 $&$ 0.0087 $\\
$ 0.500 $&$ 1.67\cdot10^{-5} $&$ 0.330 $&$ 0.4265 $&$ 0.0153 $&$ 0.0139 $&$ 0.0125 $\\
$ 0.650 $&$ 1.02\cdot10^{-3} $&$ 0.007 $&$ 0.3119 $&$ 0.0074 $&$ 0.0062 $&$ 0.0085 $\\
$ 0.650 $&$ 4.78\cdot10^{-4} $&$ 0.015 $&$ 0.3319 $&$ 0.0076 $&$ 0.0083 $&$ 0.0089 $\\
$ 0.650 $&$ 2.87\cdot10^{-4} $&$ 0.025 $&$ 0.3559 $&$ 0.0080 $&$ 0.0068 $&$ 0.0092 $\\
$ 0.650 $&$ 1.43\cdot10^{-4} $&$ 0.050 $&$ 0.3889 $&$ 0.0087 $&$ 0.0089 $&$ 0.0097 $\\
$ 0.650 $&$ 5.98\cdot10^{-5} $&$ 0.120 $&$ 0.4290 $&$ 0.0111 $&$ 0.0193 $&$ 0.0181 $\\
$ 0.650 $&$ 3.59\cdot10^{-5} $&$ 0.200 $&$ 0.4971 $&$ 0.0193 $&$ 0.0168 $&$ 0.0297 $\\
\hline \end{tabular}} \end{minipage} \normalsize

\end{sideways}

}

\caption{\label{FTWOVALUES}Measured $\ft(x,Q^2)$ values with
statistical and systematic errors. The effect of $\fl$ has been
corrected using $R$ from an approximation of the BKS model,
$R=0.165\cdot Q^2/m_\rho^2$.}{}

\end{table}

\clearpage\newpage

\clearpage\newpage

\clearpage\newpage

\clearpage\newpage

\clearpage\newpage

\clearpage\newpage

\begin{table}[p]

{\centering

\begin{sideways}

\tabcolsep2pt

{\tiny \begin{tabular}{ccrrrrrrrrrrrrrrrrrrrrrrrrrrrrr} \hline $Q^2$ &
$y$ & A$-$ & A$+$ & B$-$ & B$+$ & C$-$ & C$+$ & D$-$ & D$+$ & E\ \ \ &
F$-$ & F$+$ & G$-$ & G$+$ & H$-$ & H$+$ & I$-$ & I$+$ & J$-$ & J$+$ &
K$-$ & K$+$ & L$-$ & L$+$ & M$-$ & M$+$ & N$-$ & N$+$ & O$-$ & O$+$ \\
\hline \\

$0.045$&$0.800$&$-0.8$&$+3.0$&$ 0.0$&$ 0.0$&$-1.6$&$ 0.0$&$-0.7$&$+0.5$&$-0.4$&$+0.3$&$+2.7$&$+0.7$&$+0.4$&$-0.3$&$-1.2$&$-2.1$&$+0.5$&$-1.5$&$+2.1$&$-1.5$&$+1.5$&$-0.7$&$+0.9$&$-6.0$&$+7.8$&$-2.7$&$+1.4$&$+1.5$&$-1.5$\\
$0.065$&$0.700$&$ 0.0$&$-1.3$&$ 0.0$&$ 0.0$&$-0.4$&$-0.1$&$-1.4$&$+0.6$&$+2.4$&$+2.1$&$ 0.0$&$-0.5$&$+0.3$&$+0.1$&$+1.6$&$-0.8$&$+0.3$&$-2.5$&$+3.3$&$-1.5$&$+1.5$&$-2.1$&$+0.7$&$-3.7$&$+3.8$&$-2.1$&$+1.1$&$+2.8$&$-2.9$\\
$0.065$&$0.800$&$+1.0$&$+1.0$&$ 0.0$&$ 0.0$&$-1.7$&$+0.8$&$-1.2$&$+0.7$&$+0.8$&$+0.7$&$-2.3$&$-0.1$&$+0.1$&$+0.5$&$-0.6$&$+2.3$&$-0.9$&$-0.3$&$+0.7$&$-1.5$&$+1.5$&$-0.9$&$+0.8$&$-5.6$&$+6.3$&$-2.7$&$+1.4$&$+0.5$&$-0.5$\\
$0.085$&$0.600$&$+4.2$&$-0.6$&$ 0.0$&$ 0.0$&$-0.7$&$-0.4$&$-1.0$&$+0.7$&$-2.2$&$-1.4$&$-1.0$&$-0.6$&$+0.2$&$+1.2$&$-0.3$&$-0.2$&$-1.4$&$-1.5$&$+1.4$&$-1.5$&$+1.5$&$-0.7$&$ 0.0$&$-1.0$&$+1.0$&$-1.5$&$+0.8$&$+3.2$&$-3.2$\\
$0.085$&$0.700$&$+1.5$&$-0.9$&$ 0.0$&$ 0.0$&$-1.2$&$+0.3$&$-1.6$&$+0.8$&$+0.2$&$-1.0$&$+0.6$&$-0.6$&$+0.2$&$+0.6$&$-0.8$&$+1.3$&$-0.7$&$-0.7$&$-0.1$&$-1.5$&$+1.5$&$-0.8$&$+1.0$&$-2.9$&$+3.0$&$-2.1$&$+1.1$&$+0.6$&$-0.6$\\
$0.085$&$0.800$&$+1.5$&$ 0.0$&$ 0.0$&$ 0.0$&$-1.1$&$+1.3$&$-0.5$&$+0.7$&$+0.9$&$-0.1$&$+0.2$&$ 0.0$&$-0.1$&$+0.9$&$-0.7$&$+1.1$&$-1.5$&$-0.7$&$ 0.0$&$-1.5$&$+1.5$&$-1.3$&$+1.0$&$-4.8$&$+5.4$&$-2.7$&$+1.4$&$+0.2$&$-0.2$\\
$0.110$&$0.500$&$-0.1$&$+0.3$&$+1.4$&$-0.3$&$-0.4$&$+0.3$&$-1.0$&$+0.3$&$+1.3$&$+3.1$&$+5.5$&$-0.3$&$+0.1$&$-0.1$&$-0.5$&$-0.4$&$-0.2$&$-0.8$&$+0.5$&$-1.5$&$+1.5$&$-0.2$&$ 0.0$&$-0.3$&$+0.3$&$-0.9$&$+0.5$&$+2.6$&$-2.6$\\
$0.110$&$0.600$&$+0.4$&$-0.5$&$ 0.0$&$ 0.0$&$-0.4$&$-0.1$&$-0.7$&$+0.5$&$+0.7$&$+1.1$&$+0.1$&$-0.3$&$-0.1$&$+0.9$&$-0.7$&$+1.6$&$-1.0$&$-0.7$&$+1.3$&$-1.5$&$+1.5$&$-0.3$&$+0.5$&$-1.1$&$+1.1$&$-1.5$&$+0.8$&$+0.6$&$-0.7$\\
$0.110$&$0.700$&$+0.2$&$-0.4$&$ 0.0$&$ 0.0$&$-0.6$&$+0.2$&$-0.6$&$+0.6$&$+1.5$&$+0.4$&$+0.2$&$-1.0$&$+0.6$&$-0.1$&$-0.3$&$-0.5$&$-0.3$&$-1.5$&$+0.8$&$-1.5$&$+1.5$&$-0.7$&$+0.8$&$-2.3$&$+2.4$&$-2.1$&$+1.1$&$+0.3$&$-0.3$\\
$0.110$&$0.800$&$+0.5$&$-2.2$&$ 0.0$&$ 0.0$&$-0.6$&$+1.0$&$-1.7$&$+0.5$&$+1.3$&$+0.1$&$-0.4$&$-0.2$&$+0.3$&$+1.0$&$-0.3$&$+1.0$&$-0.4$&$-1.0$&$+1.4$&$-1.5$&$+1.5$&$-1.8$&$+1.5$&$-4.5$&$+4.9$&$-2.7$&$+1.4$&$+0.1$&$-0.2$\\
$0.150$&$0.330$&$ 0.0$&$ 0.0$&$+0.4$&$-0.2$&$+0.1$&$ 0.0$&$+0.3$&$+0.5$&$+1.8$&$-0.6$&$+1.3$&$+0.1$&$+0.4$&$+0.7$&$-0.6$&$+0.2$&$+0.6$&$-1.9$&$-0.6$&$-1.5$&$+1.5$&$-0.5$&$+0.1$&$-0.2$&$+0.2$&$ 0.0$&$ 0.0$&$+2.9$&$-2.9$\\
$0.150$&$0.400$&$ 0.0$&$ 0.0$&$+0.5$&$ 0.0$&$-0.3$&$ 0.0$&$-0.5$&$+0.3$&$-0.1$&$+0.1$&$+1.7$&$-0.1$&$+0.2$&$+1.0$&$-0.7$&$+0.5$&$-0.2$&$-0.4$&$+1.1$&$-1.5$&$+1.5$&$-0.1$&$+0.2$&$-0.1$&$+0.1$&$-0.3$&$+0.2$&$+1.1$&$-1.2$\\
$0.150$&$0.500$&$+0.1$&$-0.2$&$+0.2$&$-0.3$&$-0.3$&$+0.2$&$-0.3$&$+0.2$&$+0.6$&$+0.6$&$+0.1$&$-0.2$&$-0.6$&$+0.4$&$-0.4$&$+0.1$&$-0.6$&$-1.5$&$+1.1$&$-1.5$&$+1.5$&$-0.3$&$+0.3$&$-0.2$&$+0.2$&$-0.9$&$+0.5$&$+0.5$&$-0.6$\\
$0.150$&$0.600$&$+1.1$&$-1.1$&$ 0.0$&$ 0.0$&$-0.3$&$-0.1$&$-0.3$&$+0.5$&$+0.4$&$+0.2$&$+0.2$&$-0.6$&$-0.5$&$+0.3$&$-0.4$&$+0.6$&$-0.1$&$-0.9$&$+0.9$&$-1.5$&$+1.5$&$-0.5$&$+0.6$&$-0.8$&$+0.8$&$-1.5$&$+0.8$&$+0.4$&$-0.4$\\
$0.150$&$0.700$&$+1.4$&$-0.4$&$ 0.0$&$ 0.0$&$-0.8$&$+0.3$&$-0.3$&$+0.7$&$ 0.0$&$-0.3$&$+0.2$&$-0.6$&$-0.6$&$+0.6$&$-0.6$&$+1.3$&$-0.5$&$-0.3$&$+0.8$&$-1.5$&$+1.5$&$-0.7$&$+0.8$&$-1.7$&$+1.7$&$-2.1$&$+1.1$&$+0.2$&$-0.2$\\
$0.150$&$0.800$&$+0.5$&$-0.7$&$ 0.0$&$ 0.0$&$-0.5$&$+0.7$&$-0.6$&$+0.6$&$-0.5$&$+0.3$&$-0.6$&$+0.1$&$ 0.0$&$-0.8$&$-0.1$&$+0.3$&$-1.3$&$-0.7$&$+0.5$&$-1.5$&$+1.5$&$-1.7$&$+1.7$&$-3.6$&$+4.0$&$-2.7$&$+1.4$&$+0.1$&$-0.1$\\
$0.200$&$0.200$&$ 0.0$&$ 0.0$&$+0.5$&$-0.2$&$ 0.0$&$ 0.0$&$+0.4$&$ 0.0$&$ 0.0$&$-0.6$&$+1.1$&$-0.4$&$-0.2$&$+1.1$&$-0.7$&$-0.1$&$+0.7$&$-0.2$&$+1.2$&$-1.5$&$+1.5$&$-0.4$&$+0.2$&$-0.3$&$+0.3$&$ 0.0$&$ 0.0$&$+2.1$&$-2.2$\\
$0.200$&$0.260$&$ 0.0$&$ 0.0$&$+0.3$&$+0.1$&$ 0.0$&$ 0.0$&$+0.1$&$-0.1$&$+1.1$&$+0.1$&$-0.4$&$ 0.0$&$-0.2$&$+1.3$&$-0.7$&$+0.3$&$+0.2$&$-0.4$&$+0.5$&$-1.5$&$+1.5$&$-0.3$&$+0.2$&$ 0.0$&$ 0.0$&$ 0.0$&$ 0.0$&$+0.8$&$-0.8$\\
$0.200$&$0.330$&$ 0.0$&$ 0.0$&$ 0.0$&$-0.3$&$ 0.0$&$+0.1$&$-0.2$&$-0.2$&$-0.2$&$+0.8$&$-0.5$&$-0.5$&$-0.2$&$+0.8$&$-1.0$&$+0.4$&$-0.8$&$-0.2$&$+0.7$&$-1.5$&$+1.5$&$-0.2$&$+0.1$&$-0.1$&$+0.1$&$ 0.0$&$ 0.0$&$+0.4$&$-0.4$\\
$0.200$&$0.400$&$ 0.0$&$ 0.0$&$ 0.0$&$-0.3$&$-0.1$&$-0.1$&$+0.1$&$+0.2$&$+0.8$&$-0.2$&$-0.1$&$-0.3$&$-1.2$&$+1.1$&$-0.8$&$+1.0$&$-0.2$&$-0.9$&$+0.4$&$-1.5$&$+1.5$&$-0.3$&$+0.3$&$-0.3$&$+0.3$&$-0.3$&$+0.2$&$+0.3$&$-0.4$\\
$0.200$&$0.500$&$+0.3$&$-0.3$&$+0.4$&$+0.2$&$-0.1$&$ 0.0$&$-0.3$&$ 0.0$&$+0.3$&$+0.3$&$ 0.0$&$-0.4$&$-0.6$&$+0.8$&$-0.7$&$+0.7$&$-0.6$&$-0.6$&$+1.1$&$-1.5$&$+1.5$&$-0.2$&$+0.3$&$-0.1$&$+0.2$&$-0.9$&$+0.5$&$+0.4$&$-0.3$\\
$0.200$&$0.600$&$+0.5$&$ 0.0$&$ 0.0$&$ 0.0$&$-0.3$&$ 0.0$&$-0.6$&$+0.4$&$+0.5$&$ 0.0$&$ 0.0$&$-0.9$&$-0.7$&$+1.0$&$-0.3$&$+1.6$&$-0.6$&$-0.4$&$+0.7$&$-1.5$&$+1.5$&$-0.4$&$+0.2$&$-0.5$&$+0.5$&$-1.5$&$+0.8$&$+0.2$&$-0.2$\\
$0.200$&$0.700$&$+0.1$&$-0.3$&$ 0.0$&$ 0.0$&$-0.5$&$ 0.0$&$-1.4$&$+0.6$&$+0.3$&$-0.2$&$ 0.0$&$-1.2$&$-0.8$&$+0.6$&$-0.9$&$+0.7$&$-2.5$&$-0.6$&$+0.2$&$-1.5$&$+1.5$&$-0.5$&$+0.6$&$-1.5$&$+1.5$&$-2.1$&$+1.1$&$+0.1$&$-0.1$\\
$0.250$&$0.007$&$ 0.0$&$ 0.0$&$-0.2$&$-1.0$&$-0.1$&$-0.1$&$-0.2$&$-0.1$&$-0.5$&$+0.7$&$+5.1$&$-0.7$&$-0.2$&$+1.8$&$-2.0$&$ 0.0$&$+0.3$&$-0.9$&$+2.6$&$-1.5$&$+1.5$&$-2.0$&$-0.3$&$-0.2$&$+0.2$&$ 0.0$&$ 0.0$&$+1.5$&$-1.6$\\
$0.250$&$0.015$&$ 0.0$&$ 0.0$&$ 0.0$&$ 0.0$&$+0.1$&$ 0.0$&$+0.1$&$-0.2$&$-3.3$&$+2.6$&$+0.2$&$-0.4$&$+0.4$&$+2.3$&$-2.3$&$-0.1$&$+0.1$&$-0.1$&$-0.1$&$-1.5$&$+1.5$&$+2.1$&$+0.9$&$-0.2$&$+0.3$&$ 0.0$&$ 0.0$&$+1.4$&$-1.4$\\
$0.250$&$0.025$&$ 0.0$&$ 0.0$&$ 0.0$&$ 0.0$&$+0.1$&$-0.1$&$-0.8$&$ 0.0$&$+0.5$&$+3.6$&$+1.3$&$-0.5$&$+0.2$&$+0.6$&$-2.7$&$ 0.0$&$-1.1$&$-0.6$&$-1.3$&$-1.5$&$+1.5$&$-0.6$&$-0.7$&$-0.1$&$+0.2$&$ 0.0$&$ 0.0$&$+1.1$&$-1.2$\\
$0.250$&$0.050$&$ 0.0$&$ 0.0$&$ 0.0$&$ 0.0$&$ 0.0$&$+0.1$&$+0.4$&$+0.3$&$-0.5$&$+1.3$&$ 0.0$&$-0.7$&$+0.1$&$+1.4$&$-1.5$&$+0.1$&$-0.4$&$-2.4$&$+2.1$&$-1.5$&$+1.5$&$+1.3$&$-0.8$&$+0.3$&$-0.3$&$ 0.0$&$ 0.0$&$+0.7$&$-0.7$\\
$0.250$&$0.120$&$ 0.0$&$ 0.0$&$-0.2$&$+0.9$&$ 0.0$&$ 0.0$&$ 0.0$&$ 0.0$&$ 0.0$&$-0.2$&$+1.4$&$-0.2$&$-0.2$&$-0.2$&$-0.2$&$-0.9$&$+0.7$&$-1.0$&$+0.5$&$-1.5$&$+1.5$&$-2.3$&$+2.3$&$-0.1$&$+0.1$&$ 0.0$&$ 0.0$&$+0.2$&$-0.2$\\
$0.250$&$0.200$&$ 0.0$&$ 0.0$&$-0.2$&$+0.2$&$ 0.0$&$ 0.0$&$+0.2$&$-0.1$&$+0.4$&$+0.2$&$-0.5$&$-0.7$&$-0.4$&$+1.1$&$-1.3$&$-0.3$&$-0.2$&$-0.8$&$+1.0$&$-1.5$&$+1.5$&$-0.4$&$+0.3$&$ 0.0$&$ 0.0$&$ 0.0$&$ 0.0$&$+0.1$&$-0.1$\\
$0.250$&$0.260$&$ 0.0$&$ 0.0$&$+0.3$&$-0.2$&$ 0.0$&$ 0.0$&$ 0.0$&$+0.1$&$+0.1$&$+0.2$&$-0.2$&$-0.2$&$-0.9$&$+1.0$&$-1.0$&$+0.2$&$-0.2$&$-1.2$&$+1.2$&$-1.5$&$+1.5$&$-0.3$&$+0.3$&$-0.1$&$+0.1$&$ 0.0$&$ 0.0$&$+0.1$&$-0.1$\\
$0.250$&$0.330$&$ 0.0$&$ 0.0$&$+0.5$&$-0.4$&$ 0.0$&$ 0.0$&$+0.1$&$ 0.0$&$+0.3$&$+0.4$&$-0.6$&$-0.7$&$-1.8$&$+1.5$&$-0.9$&$+0.7$&$-0.3$&$-0.7$&$+0.6$&$-1.5$&$+1.5$&$-0.2$&$+0.3$&$ 0.0$&$ 0.0$&$ 0.0$&$ 0.0$&$+0.1$&$-0.2$\\
$0.250$&$0.400$&$ 0.0$&$ 0.0$&$+0.3$&$+0.2$&$ 0.0$&$ 0.0$&$+0.1$&$ 0.0$&$+0.6$&$+0.5$&$-0.4$&$-0.4$&$-0.7$&$+0.4$&$-1.0$&$+0.4$&$-0.7$&$-0.8$&$+0.7$&$-1.5$&$+1.5$&$-0.2$&$+0.2$&$-0.1$&$+0.1$&$-0.3$&$+0.2$&$+0.2$&$-0.3$\\
$0.250$&$0.500$&$ 0.0$&$-0.3$&$+0.1$&$+0.1$&$+0.1$&$-0.2$&$ 0.0$&$+0.4$&$+1.3$&$+0.3$&$-0.1$&$-1.0$&$-1.0$&$-0.1$&$-0.4$&$+0.4$&$-0.9$&$-0.7$&$+0.5$&$-1.5$&$+1.5$&$-0.2$&$+0.2$&$ 0.0$&$ 0.0$&$-0.9$&$+0.5$&$+0.2$&$-0.3$\\
$0.250$&$0.600$&$-0.2$&$ 0.0$&$ 0.0$&$ 0.0$&$-0.1$&$-0.2$&$-0.9$&$+0.3$&$+1.4$&$ 0.0$&$+0.4$&$+0.1$&$-1.2$&$+0.6$&$-0.9$&$+0.3$&$-0.8$&$-1.4$&$+0.4$&$-1.5$&$+1.5$&$-0.1$&$+0.2$&$-0.6$&$+0.7$&$-1.5$&$+0.8$&$+0.1$&$-0.1$\\
$0.250$&$0.700$&$+1.5$&$+0.1$&$ 0.0$&$ 0.0$&$-0.3$&$-0.8$&$-0.4$&$+0.9$&$+0.2$&$-1.6$&$-1.4$&$+0.7$&$-1.4$&$+0.1$&$+0.1$&$+0.7$&$+0.3$&$+0.1$&$+2.1$&$-1.5$&$+1.5$&$-0.9$&$+1.0$&$-1.3$&$+1.4$&$-2.1$&$+1.1$&$+0.1$&$-0.1$\\
$0.300$&$0.007$&$ 0.0$&$ 0.0$&$ 0.0$&$-1.0$&$ 0.0$&$-0.1$&$ 0.0$&$ 0.0$&$-0.5$&$+0.4$&$-0.1$&$-0.4$&$-0.6$&$+0.6$&$-0.7$&$-0.1$&$-0.2$&$-0.9$&$+0.6$&$-1.5$&$+1.5$&$-1.4$&$+1.1$&$-0.2$&$+0.2$&$ 0.0$&$ 0.0$&$ 0.0$&$-0.1$\\
$0.300$&$0.015$&$ 0.0$&$ 0.0$&$ 0.0$&$ 0.0$&$ 0.0$&$ 0.0$&$+0.7$&$+0.2$&$+1.1$&$+0.1$&$+0.9$&$-0.5$&$-0.7$&$-0.4$&$-0.3$&$ 0.0$&$ 0.0$&$-1.8$&$+1.1$&$-1.5$&$+1.5$&$+0.5$&$+0.6$&$-0.2$&$+0.2$&$ 0.0$&$ 0.0$&$ 0.0$&$ 0.0$\\
$0.300$&$0.025$&$ 0.0$&$ 0.0$&$ 0.0$&$ 0.0$&$+0.1$&$ 0.0$&$+0.4$&$ 0.0$&$+0.2$&$+0.4$&$+1.1$&$-0.1$&$-1.0$&$+0.7$&$ 0.0$&$+0.2$&$+0.1$&$+0.3$&$+1.9$&$-1.5$&$+1.5$&$-0.6$&$-0.3$&$ 0.0$&$ 0.0$&$ 0.0$&$ 0.0$&$ 0.0$&$ 0.0$\\
$0.300$&$0.050$&$ 0.0$&$ 0.0$&$ 0.0$&$ 0.0$&$ 0.0$&$ 0.0$&$-0.4$&$ 0.0$&$-0.3$&$-1.1$&$-0.2$&$-0.4$&$-1.7$&$+0.3$&$+0.4$&$-0.1$&$+0.3$&$-0.6$&$+1.2$&$-1.5$&$+1.5$&$+0.2$&$-0.3$&$+0.1$&$-0.2$&$ 0.0$&$ 0.0$&$ 0.0$&$ 0.0$\\
$0.300$&$0.120$&$ 0.0$&$ 0.0$&$+0.5$&$+0.4$&$ 0.0$&$ 0.0$&$+0.1$&$ 0.0$&$-0.1$&$-0.2$&$-0.2$&$-0.8$&$-1.1$&$-0.4$&$+0.1$&$-1.1$&$+1.1$&$-1.0$&$+1.2$&$-1.5$&$+1.5$&$-2.2$&$+2.5$&$ 0.0$&$-0.1$&$ 0.0$&$ 0.0$&$ 0.0$&$ 0.0$\\
$0.300$&$0.200$&$ 0.0$&$ 0.0$&$+0.1$&$ 0.0$&$ 0.0$&$ 0.0$&$+0.1$&$+0.1$&$+0.6$&$+0.3$&$-0.2$&$-0.4$&$-1.7$&$+2.0$&$-1.1$&$+0.8$&$ 0.0$&$-0.7$&$+0.6$&$-1.5$&$+1.5$&$-0.4$&$+0.3$&$-0.1$&$+0.1$&$ 0.0$&$ 0.0$&$ 0.0$&$ 0.0$\\
$0.300$&$0.260$&$ 0.0$&$ 0.0$&$+0.2$&$-0.1$&$+0.1$&$ 0.0$&$-0.1$&$-0.1$&$+0.8$&$+0.2$&$-0.1$&$-1.2$&$-0.8$&$+0.9$&$-1.4$&$ 0.0$&$-0.3$&$-0.6$&$+0.9$&$-1.5$&$+1.5$&$-0.2$&$+0.2$&$-0.1$&$+0.1$&$ 0.0$&$ 0.0$&$ 0.0$&$ 0.0$\\
$0.300$&$0.330$&$ 0.0$&$ 0.0$&$+0.3$&$-0.2$&$+0.2$&$ 0.0$&$+0.2$&$+0.1$&$+0.6$&$+0.1$&$-0.1$&$-1.2$&$-1.3$&$ 0.0$&$-1.0$&$+0.1$&$-0.7$&$-0.9$&$+0.8$&$-1.5$&$+1.5$&$-0.2$&$+0.3$&$-0.1$&$+0.1$&$ 0.0$&$ 0.0$&$ 0.0$&$ 0.0$\\
$0.300$&$0.400$&$ 0.0$&$ 0.0$&$+0.1$&$-0.4$&$+0.2$&$ 0.0$&$-0.2$&$+0.1$&$+0.7$&$+0.1$&$+0.1$&$-0.7$&$-0.7$&$+1.5$&$-0.7$&$+1.0$&$-0.1$&$-1.1$&$+1.2$&$-1.5$&$+1.5$&$-0.2$&$+0.1$&$-0.1$&$+0.1$&$-0.3$&$+0.2$&$+0.1$&$-0.1$\\
$0.300$&$0.500$&$ 0.0$&$+0.1$&$+0.5$&$+0.3$&$ 0.0$&$-0.2$&$-0.5$&$+0.2$&$+0.4$&$-0.6$&$-0.2$&$-0.6$&$-0.2$&$-0.2$&$-0.1$&$-0.6$&$-0.3$&$-1.3$&$+0.8$&$-1.5$&$+1.5$&$-0.3$&$+0.3$&$-0.2$&$+0.3$&$-0.9$&$+0.5$&$+0.1$&$-0.1$\\
$0.300$&$0.600$&$-1.4$&$-0.4$&$ 0.0$&$ 0.0$&$+0.5$&$+0.2$&$-0.3$&$ 0.0$&$+3.4$&$-1.1$&$-0.2$&$+0.3$&$-2.4$&$+1.6$&$-0.3$&$+1.2$&$-0.3$&$+0.2$&$+1.7$&$-1.5$&$+1.5$&$-0.3$&$+0.2$&$-0.5$&$+0.7$&$-1.5$&$+0.8$&$+0.1$&$-0.1$\\
$0.400$&$0.007$&$ 0.0$&$ 0.0$&$ 0.0$&$+0.7$&$ 0.0$&$ 0.0$&$-0.5$&$+0.2$&$+1.5$&$+0.5$&$ 0.0$&$-0.7$&$-1.5$&$-0.9$&$ 0.0$&$-0.3$&$+0.1$&$-1.2$&$+0.8$&$-1.5$&$+1.5$&$-0.7$&$ 0.0$&$-0.2$&$+0.3$&$ 0.0$&$ 0.0$&$ 0.0$&$ 0.0$\\
$0.400$&$0.015$&$ 0.0$&$ 0.0$&$ 0.0$&$ 0.0$&$ 0.0$&$ 0.0$&$+0.1$&$-0.1$&$+0.7$&$ 0.0$&$-0.2$&$-0.4$&$-1.4$&$+0.3$&$+0.8$&$ 0.0$&$ 0.0$&$-0.4$&$+1.2$&$-1.5$&$+1.5$&$+0.3$&$+0.9$&$-0.1$&$+0.1$&$ 0.0$&$ 0.0$&$ 0.0$&$ 0.0$\\
$0.400$&$0.025$&$ 0.0$&$ 0.0$&$ 0.0$&$ 0.0$&$ 0.0$&$+0.1$&$+0.2$&$-0.2$&$+1.1$&$+0.2$&$-0.1$&$-1.5$&$-1.7$&$-0.4$&$ 0.0$&$ 0.0$&$-0.1$&$-0.8$&$+1.1$&$-1.5$&$+1.5$&$+0.8$&$+0.1$&$ 0.0$&$+0.1$&$ 0.0$&$ 0.0$&$ 0.0$&$ 0.0$\\
$0.400$&$0.050$&$ 0.0$&$ 0.0$&$ 0.0$&$ 0.0$&$ 0.0$&$-0.1$&$+0.1$&$-0.2$&$+0.7$&$+0.1$&$-0.2$&$-2.0$&$-2.0$&$-0.5$&$-0.9$&$+0.1$&$+0.4$&$-1.2$&$-0.2$&$-1.5$&$+1.5$&$-1.6$&$-0.7$&$+0.2$&$-0.2$&$ 0.0$&$ 0.0$&$ 0.0$&$ 0.0$\\
$0.400$&$0.120$&$ 0.0$&$ 0.0$&$+0.2$&$+0.3$&$ 0.0$&$ 0.0$&$+0.3$&$ 0.0$&$+0.3$&$ 0.0$&$-0.1$&$-0.4$&$-1.5$&$ 0.0$&$-0.4$&$-1.0$&$+0.7$&$-0.8$&$+0.4$&$-1.5$&$+1.5$&$-2.6$&$+2.5$&$ 0.0$&$ 0.0$&$ 0.0$&$ 0.0$&$ 0.0$&$ 0.0$\\
$0.400$&$0.200$&$ 0.0$&$ 0.0$&$-0.4$&$+0.7$&$+0.1$&$-0.1$&$+0.3$&$-0.1$&$+0.9$&$-0.1$&$-0.1$&$-0.9$&$-1.3$&$+0.6$&$-0.8$&$-0.6$&$+0.4$&$-0.7$&$+1.1$&$-1.5$&$+1.5$&$-0.3$&$+0.4$&$+0.2$&$-0.1$&$ 0.0$&$ 0.0$&$ 0.0$&$ 0.0$\\
$0.400$&$0.260$&$ 0.0$&$ 0.0$&$+0.1$&$-0.2$&$ 0.0$&$+0.1$&$+0.1$&$ 0.0$&$+1.1$&$ 0.0$&$ 0.0$&$-0.5$&$-0.9$&$+1.1$&$-0.4$&$+0.5$&$+0.3$&$-1.0$&$+1.2$&$-1.5$&$+1.5$&$-0.1$&$+0.2$&$-0.1$&$+0.1$&$ 0.0$&$ 0.0$&$ 0.0$&$ 0.0$\\
$0.400$&$0.330$&$ 0.0$&$ 0.0$&$+0.5$&$-0.5$&$-0.1$&$-0.1$&$+0.3$&$+0.1$&$+1.9$&$+0.1$&$-0.2$&$-0.8$&$-1.7$&$+0.3$&$-0.7$&$-0.9$&$ 0.0$&$-1.3$&$+1.0$&$-1.5$&$+1.5$&$-0.3$&$+0.3$&$+0.2$&$-0.2$&$ 0.0$&$ 0.0$&$ 0.0$&$ 0.0$\\
$0.400$&$0.400$&$ 0.0$&$ 0.0$&$+0.5$&$+0.3$&$+0.2$&$-0.1$&$-0.1$&$+0.1$&$+2.1$&$+0.4$&$-0.3$&$+0.1$&$-0.9$&$+1.3$&$-1.2$&$+1.4$&$-0.9$&$-0.3$&$+0.6$&$-1.5$&$+1.5$&$-0.2$&$ 0.0$&$-0.2$&$+0.3$&$-0.3$&$+0.2$&$ 0.0$&$ 0.0$\\
$0.400$&$0.500$&$-0.5$&$+0.4$&$-0.5$&$-0.2$&$+0.6$&$+0.1$&$-0.7$&$+0.1$&$+2.7$&$+0.9$&$+0.2$&$+0.6$&$-0.8$&$+1.4$&$-0.1$&$+1.5$&$+0.8$&$+1.1$&$+0.3$&$-1.5$&$+1.5$&$-0.1$&$+0.1$&$-0.2$&$+0.3$&$-0.9$&$+0.5$&$ 0.0$&$ 0.0$\\
$0.500$&$0.007$&$ 0.0$&$ 0.0$&$ 0.0$&$-0.2$&$ 0.0$&$-0.1$&$+0.2$&$-0.1$&$-0.3$&$ 0.0$&$+0.1$&$-0.5$&$-0.7$&$+0.9$&$-0.8$&$ 0.0$&$+0.1$&$-0.9$&$+0.7$&$-1.5$&$+1.5$&$-0.5$&$+0.4$&$-0.1$&$+0.2$&$ 0.0$&$ 0.0$&$ 0.0$&$ 0.0$\\
$0.500$&$0.015$&$ 0.0$&$ 0.0$&$ 0.0$&$ 0.0$&$ 0.0$&$ 0.0$&$ 0.0$&$-0.1$&$-0.4$&$-0.1$&$+0.1$&$ 0.0$&$-0.6$&$+0.6$&$-0.9$&$+0.1$&$+0.1$&$-1.0$&$+0.7$&$-1.5$&$+1.5$&$+0.2$&$+0.5$&$-0.2$&$+0.2$&$ 0.0$&$ 0.0$&$ 0.0$&$ 0.0$\\
$0.500$&$0.025$&$ 0.0$&$ 0.0$&$ 0.0$&$ 0.0$&$+0.1$&$ 0.0$&$+0.4$&$ 0.0$&$-0.3$&$ 0.0$&$+0.2$&$-0.9$&$-1.0$&$+0.3$&$-0.2$&$-0.1$&$+0.2$&$-0.8$&$+1.4$&$-1.5$&$+1.5$&$ 0.0$&$-1.0$&$-0.2$&$+0.2$&$ 0.0$&$ 0.0$&$ 0.0$&$ 0.0$\\
$0.500$&$0.050$&$ 0.0$&$ 0.0$&$ 0.0$&$ 0.0$&$+0.1$&$ 0.0$&$+0.3$&$ 0.0$&$+0.1$&$ 0.0$&$+0.1$&$-0.6$&$-1.0$&$+0.2$&$-0.5$&$-0.2$&$+0.1$&$-1.1$&$+0.9$&$-1.5$&$+1.5$&$+0.4$&$+0.1$&$+0.2$&$-0.2$&$ 0.0$&$ 0.0$&$ 0.0$&$ 0.0$\\
$0.500$&$0.120$&$ 0.0$&$ 0.0$&$-0.4$&$-1.3$&$+0.1$&$ 0.0$&$+0.3$&$-0.1$&$+1.4$&$ 0.0$&$+0.1$&$-1.0$&$-1.8$&$+0.5$&$+0.7$&$-0.7$&$+1.3$&$-0.6$&$+1.1$&$-1.5$&$+1.5$&$-2.1$&$+2.6$&$-0.1$&$+0.1$&$ 0.0$&$ 0.0$&$ 0.0$&$ 0.0$\\
$0.500$&$0.200$&$ 0.0$&$ 0.0$&$+0.1$&$+0.5$&$ 0.0$&$ 0.0$&$ 0.0$&$-0.1$&$+1.7$&$-0.2$&$ 0.0$&$-0.9$&$-1.6$&$+1.0$&$-1.5$&$-0.2$&$-0.3$&$-1.1$&$+0.7$&$-1.5$&$+1.5$&$-0.5$&$+0.3$&$-0.2$&$+0.3$&$ 0.0$&$ 0.0$&$ 0.0$&$ 0.0$\\
$0.500$&$0.260$&$ 0.0$&$ 0.0$&$ 0.0$&$-0.1$&$+0.1$&$ 0.0$&$+0.7$&$-0.1$&$+1.7$&$+0.9$&$-0.4$&$-0.2$&$-1.0$&$+1.0$&$-1.3$&$+0.2$&$-0.2$&$-0.4$&$ 0.0$&$-1.5$&$+1.5$&$-0.2$&$+0.2$&$-0.1$&$+0.1$&$ 0.0$&$ 0.0$&$ 0.0$&$ 0.0$\\
$0.500$&$0.330$&$ 0.0$&$ 0.0$&$-0.4$&$+0.4$&$-0.1$&$-0.1$&$-0.2$&$-0.1$&$+2.6$&$-0.4$&$-1.7$&$+0.5$&$-1.1$&$+0.9$&$-1.0$&$+0.6$&$-1.0$&$+0.4$&$+0.1$&$-1.5$&$+1.5$&$-0.2$&$ 0.0$&$-0.1$&$+0.1$&$ 0.0$&$ 0.0$&$ 0.0$&$ 0.0$\\
$0.650$&$0.007$&$ 0.0$&$ 0.0$&$ 0.0$&$-0.4$&$ 0.0$&$+0.1$&$ 0.0$&$ 0.0$&$-0.2$&$-0.7$&$+0.2$&$-0.2$&$-1.8$&$-0.8$&$+0.8$&$-0.1$&$ 0.0$&$-0.7$&$+0.8$&$-1.5$&$+1.5$&$-0.2$&$+0.2$&$-0.4$&$+0.4$&$ 0.0$&$ 0.0$&$ 0.0$&$ 0.0$\\
$0.650$&$0.015$&$ 0.0$&$ 0.0$&$ 0.0$&$ 0.0$&$+0.1$&$ 0.0$&$+0.4$&$-0.1$&$+0.2$&$-0.7$&$-0.2$&$-0.4$&$-1.7$&$-0.7$&$+1.2$&$-0.1$&$+0.2$&$-0.9$&$+1.2$&$-1.5$&$+1.5$&$+0.4$&$+0.9$&$-0.2$&$+0.2$&$ 0.0$&$ 0.0$&$ 0.0$&$ 0.0$\\
$0.650$&$0.025$&$ 0.0$&$ 0.0$&$ 0.0$&$ 0.0$&$+0.1$&$ 0.0$&$+0.3$&$-0.1$&$+0.1$&$+0.7$&$+0.2$&$ 0.0$&$-1.7$&$-0.9$&$+0.6$&$-0.3$&$+0.1$&$-0.4$&$+0.4$&$-1.5$&$+1.5$&$-0.8$&$+0.3$&$-0.2$&$+0.2$&$ 0.0$&$ 0.0$&$ 0.0$&$ 0.0$\\
$0.650$&$0.050$&$ 0.0$&$ 0.0$&$ 0.0$&$ 0.0$&$+0.1$&$ 0.0$&$+0.1$&$-0.2$&$+0.4$&$+1.0$&$-0.7$&$+0.1$&$-0.9$&$-1.2$&$+1.1$&$-0.4$&$+0.3$&$-0.9$&$+0.7$&$-1.5$&$+1.5$&$-0.4$&$ 0.0$&$+0.2$&$-0.3$&$ 0.0$&$ 0.0$&$ 0.0$&$ 0.0$\\
$0.650$&$0.120$&$ 0.0$&$ 0.0$&$-1.6$&$-0.5$&$+0.2$&$ 0.0$&$+0.2$&$-0.1$&$+3.2$&$+0.9$&$-0.8$&$+0.7$&$-1.8$&$-0.8$&$+0.1$&$-1.3$&$+0.6$&$-0.3$&$ 0.0$&$-1.5$&$+1.5$&$-2.6$&$+2.5$&$-0.1$&$+0.1$&$ 0.0$&$ 0.0$&$ 0.0$&$ 0.0$\\
$0.650$&$0.200$&$ 0.0$&$ 0.0$&$-1.3$&$-0.8$&$-0.1$&$-0.1$&$-0.5$&$+0.2$&$+2.2$&$+0.4$&$-4.9$&$+0.4$&$-1.2$&$+1.9$&$-2.3$&$+0.6$&$-0.1$&$+0.3$&$-0.6$&$-1.5$&$+1.5$&$-0.5$&$+0.2$&$-0.1$&$+0.2$&$ 0.0$&$ 0.0$&$ 0.0$&$ 0.0$\\

\hline \end{tabular}}

\end{sideways}

}

\hfill

Table 2: Individual effects on $\ft$ of each of the systematic checks
described in Section \ref{sec:SYS} (in percent). The letters denote
variations of: A) $\delta$ cut; B) $y_\mathrm{JB}$ cut; C) BPC shower
width cut; D) BPC/BPT track match cut; E) BPT vertex cut; F) fiducial
area cut in $X$; G) fiducial area cut in $Y$; H) BPC energy scale; I)
BPC energy linearity; J) BPC/BPT alignment; K) BPT efficiency; L) CAL
energy scale; M) diffractive fraction; N) background normalization; O)
radiative corrections.

\end{table} 

\end{document}